\documentclass[manuscript,screen]{acmart}

\usepackage{amsmath,amsfonts}
\usepackage{array}
\usepackage{diagbox}
\usepackage{booktabs} 
\usepackage{colortbl}   
\usepackage{multirow}
\usepackage{makecell}
\usepackage{tablefootnote}
\usepackage{arydshln}
\usepackage{tikz}
\usepackage{longtable}
\usepackage{makecell}
\usepackage{pifont}
\usepackage{acronym}
\usepackage{fancyvrb}
\usepackage{tabularx}
\usepackage{subcaption}

\usepackage{algorithm}
\usepackage{algpseudocode}

\newcommand{\std}[1]{\scriptsize{(#1)}\normalsize}

\newcommand{\thickhline}{\noalign{\hrule height 0.7pt}}

\acrodef{LLM}{large language model}
\acrodef{IR}{information retrieval}

\AtBeginDocument{%
  }

\setcopyright{acmlicensed}
\acmJournal{TOIS}
\copyrightyear{2026}
\acmYear{2026}
\acmDOI{}
\acmISBN{}

\begin{document}

\title{On the Robustness of LLM-Based Dense Retrievers: A Systematic Analysis of Generalizability and Stability}


\author{Yongkang Li}
\email{y.li7@uva.nl}
\orcid{0000-0001-6837-6184}
\affiliation{%
  \institution{University of Amsterdam}
  \city{Amsterdam}
  \country{The Netherlands}
}

\author{Panagiotis Eustratiadis}
\orcid{0000-0002-9407-1293}
\email{p.efstratiadis@uva.nl}
\affiliation{%
  \institution{University of Amsterdam}
  \city{Amsterdam}
  \country{The Netherlands}
}
\author{Yixing Fan}
\orcid{0000-0003-4317-2702}
\email{fanyixing@ict.ac.cn}
\affiliation{%
  \institution{Chinese Academy of Sciences}
  \city{Beijing}
  \country{China}
}

\author{Evangelos Kanoulas}
\orcid{0000-0002-8312-0694}
\email{e.kanoulas@uva.nl}
\affiliation{%
  \institution{University of Amsterdam}
  \city{Amsterdam}
  \country{The Netherlands}
}

\begin{abstract}
Decoder-only large language models (LLMs) are increasingly replacing BERT-style architectures as the backbone for dense retrieval, achieving substantial performance gains and broad adoption. However, the robustness of these LLM-based retrievers remains underexplored. In this paper, we present the first systematic study of the robustness of state-of-the-art open-source LLM-based dense retrievers from two complementary perspectives: \textit{generalizability} and \textit{stability}. For generalizability, we evaluate retrieval effectiveness across four benchmarks spanning 30 datasets, using linear mixed-effects models to estimate marginal mean performance and disentangle intrinsic model capability from dataset heterogeneity.
Our analysis reveals that while instruction-tuned models generally excel, those optimized for complex reasoning often suffer a ``specialization tax,'' exhibiting limited generalizability in broader contexts. For stability, we assess model resilience against both unintentional query variations~(e.g., paraphrasing, typos) and malicious adversarial attacks~(e.g., corpus poisoning). We find that LLM-based retrievers show improved robustness against typos and corpus poisoning compared to encoder-only baselines, yet remain vulnerable to semantic perturbations like synonymizing. Further analysis shows that embedding geometry (e.g., angular uniformity) provides predictive signals for lexical stability and suggests that scaling model size generally improves robustness. These findings inform future robustness-aware retriever design and principled benchmarking. Our code is publicly available at \url{https://github.com/liyongkang123/Robust_LLM_Retriever_Eval}.
\end{abstract}



\received{20 February 2007}
\received[revised]{12 March 2009}
\received[accepted]{5 June 2009}

\maketitle

\section{Introduction}

\Acf{IR} is fundamental to how humans access information in the digital age.
Recently, dense retrieval has established itself as a leading paradigm, distinguished by its ability to capture semantic nuances while maintaining efficiency~\cite{KarpukhinOMLWEC20_DPR,douze2024faiss}.
In this paradigm, queries and documents are encoded into high-dimensional vectors, and relevance is computed via distance functions such as inner product or cosine similarity.

Early dense retrievers predominantly employed BERT-style~\cite{DevlinCLT19_bert} encoder-only architectures typically trained using contrastive learning or distillation-based objectives, with representative examples including DPR~\cite{KarpukhinOMLWEC20_DPR}, Contriever~\cite{izacard2021contriever}, TAS-B~\cite{HofstatterLYLH21_tasb_dense_retrieval}, and DRAGON~\cite{LinALOLMY023_dragon}. With the rise of \acp{LLM}, decoder-only architectures like LLaMA~\cite{touvron2023llama}, Mistral~\cite{Mistral_7B}, and Qwen~\cite{bai2023qwen} have prompted researchers to explore their potential as dense encoders. This progression spans from initial attempts like RepLLaMA~\cite{MaWYWL24_RankLLaMA} and LLM2Vec~\cite{li-etal-2024-llama2vec} to current state-of-the-art~(SOTA) models like Qwen3-Embedding~\cite{qwen3embedding} and DIVER~\cite{long2025diver}. Consequently, these \ac{LLM}-based retrievers now dominate major authoritative benchmarks, such as BEIR~\cite{thakur2021beir} and BRIGHT~\cite{su2024bright}.

Despite this remarkable progress, the robustness of these LLM-based retrievers remains largely unexplored. While leaderboard metrics quantify performance on standard test sets, they often fail to reflect reliability in dynamic, open-world environments. This paper addresses that gap by conducting the first systematic analysis of the robustness of SOTA open-source decoder-only LLM-based dense retrievers. Drawing on the framing of \citet{WuZGFC23} and \citet{liu2024robust}, we study two complementary aspects of robustness: \textit{generalizability}, which concerns model performance across diverse evaluation scenarios, and \textit{stability}, which concerns model resilience under perturbations. We further examine which intrinsic model properties are associated with robustness, aiming to provide guidance for the evaluation and design of more reliable retrievers.

Specifically, our study is guided by the following three research questions:

\textbf{RQ1: How generalizable are LLM-based dense retrievers across diverse conditions?}
We address this by evaluating SOTA open-source LLM-based retrievers on a comprehensive suite of 30 datasets spanning 4 benchmarks: MS MARCO~\cite{bajaj2016ms}, BEIR~\cite{thakur2021beir}, BRIGHT~\cite{su2024bright}, and BrowseComp-Plus~\cite{chen2025browsecomp}.
To enable fine-grained analysis, we categorize these datasets according to 11 task types, 8 query types, and 5 corpus source types.
Critically, standard aggregation metrics (e.g., macro-averaging) can be misleading due to significant variances in dataset size and query difficulty.
To overcome this, we introduce \textit{linear mixed-effects models (LMMs)}~\cite{Gałecki2013_LMM} to estimate marginal mean performance while controlling for dataset-specific random effects, enabling a fairer assessment of intrinsic model capabilities.
Our analysis reveals that instruction-augmented LLM-based retrievers achieve the strongest average estimated performance across all three factors.
However, models explicitly optimized for complex reasoning-intensive tasks often suffer from a ``specialization tax'', exhibiting limited generalizability across broader retrieval conditions.

\textbf{RQ2: How stable are LLM-based dense retrievers under perturbations?}
We assess stability from two complementary angles: query-side variations and document-side adversarial attacks.
For query-side stability, we consider five query variations---misspelling, reordering, synonymizing, paraphrasing, and naturalizing---following prior work~\cite{PenhaCH22_query_variation}, and measure the relative performance degradation under each type.
For document-side stability, we evaluate resilience against corpus poisoning attacks~\cite{ZhongHWC23_Poisoning,li2025reproducinghotflip} under both white-box and direct-transfer black-box settings, where adversarial passages are injected to mislead the retriever.
Our experiments reveal that while LLM-based retrievers demonstrate improved robustness against misspellings and white-box corpus poisoning over encoder-only models, synonymizing remains broadly challenging across model families, whereas paraphrasing is more dataset-dependent and especially disruptive in shorter-query settings. By contrast, direct-transfer corpus poisoning proves largely ineffective in the no-access setting we study.

\textbf{RQ3: What factors are predictive of robustness in LLM-based retrievers?}
We investigate potential predictors of robustness, focusing on the geometry of embedding spaces---specifically isotropy---as well as model size and spectral norm.
Our analyses suggest that different factors relate to different facets of stability. \textit{Angular uniformity} (proxied by average pairwise cosine similarity) provides the clearest cross-model signal of vulnerability to lexical and other surface-level perturbations, whereas \textit{variance uniformity} (proxied by IsoScore) shows only suggestive, context-dependent trends. Beyond isotropy, larger models within the Qwen3 family tend to be more robust, while spectral norm shows limited predictive value in our setting. Two controlled interventions---angular uniformity regularization and bidirectional attention conversion---further show no consistent robustness gains in our intervention setting. Together, these results suggest that angular uniformity is more useful as a diagnostic signal than as a standalone training target, while attention directionality remains a plausible but unverified contributor.

Our main contributions are as follows:

$\bullet$ To our knowledge, we present the first systematic robustness study of decoder-only LLM-based dense retrievers, synthesizing a two-dimensional framework---generalizability and stability---for comprehensive evaluation.

$\bullet$  We propose a refined evaluation protocol integrating linear mixed-effects models to isolate intrinsic model performance from dataset variances, enabling fairer and statistically rigorous comparisons.\looseness=-1

$\bullet$ We show that robustness gains in LLM-based retrievers are perturbation-specific: instruction-augmented models are more robust to character-level noise and white-box corpus poisoning, yet synonymizing remains broadly challenging and paraphrasing is more dataset-dependent. We further identify embedding isotropy as the clearest cross-model signal of lexical and surface-level robustness.

$\bullet$ We conduct controlled intervention experiments to probe the geometry--robustness boundary and synthesize the findings into concrete guidelines for robustness-aware evaluation, evidence-based model selection, and training objective design targeting query reformulation invariance.

\section{Related Work}

We organize related work along three axes: retrieval paradigms (Section~\ref{sec:rw_models}), robustness evaluation in neural retrieval (Section~\ref{sec:rw_robustness}), and embedding geometry as a lens for analyzing robustness (Section~\ref{sec:rw_geometry}).

\subsection{Retrieval Paradigms and the Rise of LLM-based Dense Retrieval}
\label{sec:rw_models}

We organize prior retrieval models into three broad paradigms: sparse lexical retrieval, single-vector dense retrieval, and multi-vector late interaction. This taxonomy situates contemporary LLM-based dense retrievers within the broader evolution of retrieval architectures.

\subsubsection{Sparse Retrieval}

Information retrieval was long dominated by sparse lexical methods. Classical models such as TF-IDF and BM25~\cite{RobertsonZ09_BM25} represent queries and documents as high-dimensional sparse vectors and compute relevance through exact term matching. These methods remain strong baselines because of their efficiency and interpretability, but they suffer from the vocabulary mismatch problem: semantically related queries and documents with limited lexical overlap are poorly matched. Learned sparse models such as SPLADE~\cite{FormalPC21_splade} partially mitigate this limitation by combining sparse representations with neural term weighting. More broadly, the limitations of exact matching motivated the shift toward dense semantic retrieval.

\subsubsection{Neural Dense Retrieval: From Encoder-only to LLM-based Models}

Pre-trained language models, especially BERT~\cite{DevlinCLT19_bert}, enabled the modern dense retrieval paradigm, in which bi-encoders map queries and documents into a shared embedding space and compute relevance via dot product or cosine similarity. DPR~\cite{KarpukhinOMLWEC20_DPR} established the effectiveness of contrastive fine-tuning, and subsequent work improved the paradigm through stronger pre-training~\cite{XiaoLSC22_RetroMAE,izacard2021contriever}, knowledge distillation~\cite{HofstatterLYLH21_tasb_dense_retrieval,LinALOLMY023_dragon}, and multilingual modeling~\cite{chen-etal-2024-m3}. With the rise of decoder-only \acp{LLM}---including LLaMA~\cite{touvron2023llama}, Mistral~\cite{Mistral_7B}, and Qwen~\cite{bai2023qwen}---researchers began adapting them into dense encoders through various pooling strategies, typically last-token pooling~\cite{MaWYWL24_RankLLaMA,li-etal-2024-llama2vec}. A later generation of instruction-augmented retrievers, including Linq~\cite{choi2024linq}, GTE~\cite{li2023towards}, NV-Retriever~\cite{moreira2024nv}, and Qwen3-Embedding~\cite{qwen3embedding}, further improved transfer across retrieval tasks and achieved strong performance on broad evaluation benchmarks such as BEIR~\cite{thakur2021beir}.

More recently, a further specialization has emerged in retrieval models designed for reasoning-intensive queries. ReasonIR~\cite{shao2025reasonir} incorporates chain-of-thought traces into training, DIVER~\cite{long2025diver} adopts a multi-stage training scheme for complex inference, and ReasonEmbed~\cite{chen2025reasonembedenhancedtextembeddings} augments the Qwen3 backbone with logical dependency modeling. These models perform strongly on reasoning-oriented benchmarks such as BRIGHT~\cite{su2024bright}, but their growing specialization also raises a broader question: whether improvements on reasoning-heavy benchmarks translate into robust performance across more diverse retrieval conditions.

\subsubsection{Multi-vector and Late Interaction Models}

Beyond single-vector bi-encoders, another line of work represents queries and documents as sets of token-level embeddings and computes relevance through late interaction. ColBERT~\cite{KhattabZ20_ColBERT} introduced the MaxSim operator, which offers a favorable trade-off between expressiveness and efficiency. Later extensions, including ColBERTv2~\cite{SanthanamKSPZ22_ColBERTv2} and PLAID~\cite{SanthanamKPZ22_PLAID}, further improved compression and retrieval efficiency.

While these methods are compelling alternatives, they operate under a different retrieval paradigm. Recent theoretical work has also highlighted limitations of single-embedding dense retrieval~\cite{weller2025theoretical}. The present study therefore focuses on single-vector LLM-based dense retrievers, which dominate current leaderboards and are widely deployed, leaving the robustness analysis of multi-vector and sparse-neural systems to future work.

\subsection{Robustness in Neural Retrieval}
\label{sec:rw_robustness}

Prior work on retrieval robustness can be broadly grouped into two lines: generalization across domains and tasks, and stability under perturbations or adversarial manipulation. We review both lines of work before briefly discussing evidence from retrieval-adjacent components.

\subsubsection{Generalizability and Out-of-Distribution Evaluation}

A central challenge for dense retrievers is generalizability: whether models trained on one domain or task transfer reliably to unseen distributions. BEIR~\cite{thakur2021beir} was instrumental in exposing this weakness, showing that early BERT-based dense retrievers often underperform BM25~\cite{RobertsonZ09_BM25} in zero-shot out-of-distribution~(OOD) evaluation. Related model-centric work has sought to mitigate these weaknesses by improving lexical matching and OOD transfer within dense retrieval itself~\cite{chen-etal-2022-salient}. BRIGHT~\cite{su2024bright} later introduced a more demanding reasoning-oriented benchmark, where the gap between lexical and neural methods narrows substantially and even strong LLM-based retrievers remain challenged.

Broader conceptual frameworks have been proposed by \citet{WuZGFC23} and \citet{liu2024robust}, who characterize retrieval robustness in terms of performance variance, OOD generalization, and adversarial resilience. Yet the empirical literature remains fragmented: some studies examine benchmark-level transfer, others focus on adversarial vulnerability, and most center on encoder-based retrievers. Consequently, how modern LLM-based dense retrievers generalize across heterogeneous tasks, query types, and corpus sources remains insufficiently understood, particularly under statistically unbalanced evaluation settings.

\subsubsection{Stability Under Perturbations and Adversarial Attacks}

Within this line of work, prior studies have mainly examined three settings: query-side variation, document-side adversarial manipulation, and robustness issues in retrieval-adjacent components.

On the query side, real-world queries are often noisy and variable: users make typographical errors, rephrase requests, and express the same information need in multiple ways. Prior work has shown that such variation can substantially affect retrieval quality. \citet{zhuang2021dealing} and \citet{CharacterBERT} showed that BERT-based dense retrievers are sensitive to character-level noise such as misspellings, and similar vulnerabilities have also been reported for dual-encoder dense retrievers more broadly~\cite{10.1145/3477495.3531818}. Subsequent work further explored how to improve typo robustness in dense retrievers~\cite{SidiropoulosK24}. \citet{PenhaCH22_query_variation} systematically studied five types of query variation---misspelling, reordering, synonymizing, paraphrasing, and naturalizing---and showed that even semantically equivalent reformulations can cause sizable performance drops. Most closely related, \citet{hagen-etal-2024-revisiting} revisited query variation robustness for a set of neural retrievers including early LLM-based models, but the evaluation remains limited to query-side perturbations on relatively small-scale benchmarks. The robustness of current LLM-based retrievers under a broader range of query conditions therefore remains insufficiently characterized.

Beyond query-side perturbations, retrieval systems are also vulnerable to adversarial corpus manipulation. Corpus poisoning attacks~\cite{ZhongHWC23_Poisoning} inject carefully crafted passages into the document index in order to hijack retrieval results for target queries. Gradient-based corpus poisoning methods optimize adversarial passages against target query embeddings. Token-level approaches such as HotFlip search for adversarial tokens in discrete space~\cite{li2025reproducinghotflip,su2024corpus}, while \citet{li2025unsupervised} operate directly in continuous embedding space without discrete token search. \citet{10.1145/3583780.3614793_MCARA} proposed a black-box alternative based on multi-view contrastive learning. Related security work has also studied black-box adversarial attacks against neural ranking systems, for example PRADA and Order-Disorder~\cite{10.1145/3576923_PRADA,10.1145/3548606.3560683_Order-Disorder}. From a complementary security perspective, \citet{Ben-TovS25} study vulnerabilities in dense embedding-based search more broadly. Although these attacks have been studied extensively for BERT-based retrievers, their effectiveness against LLM-based encoders remains unclear, and comparative evidence across modern LLM-based retrievers is still limited.

Adversarial vulnerabilities have also been observed in adjacent components of the retrieval pipeline. \citet{tamber2025illusions} showed that LLM-based relevance judges can be misled by document-level content injection, while \citet{BehnamGhaderMR25} examined retriever behavior under malicious queries. These findings motivate robustness evaluation across the full retrieval pipeline, while also underscoring the need to isolate the retrieval backbone as a central object of analysis.

Taken together, prior work establishes robustness as a multi-dimensional property of retrieval systems, but the evidence remains incomplete in three respects. First, generalizability studies mostly emphasize benchmark-level transfer rather than factor-level variation across tasks, query intents, and corpus sources. Second, stability studies typically isolate query-side perturbations from document-side attacks rather than evaluating both within a unified setup. Third, studies centered specifically on modern LLM-based dense retrievers, including recent work on robustness scaling~\cite{Liu0GRFC25}, remain limited in scope. This leaves a clear need for a systematic robustness analysis of contemporary LLM-based retrievers within a broader retrieval robustness framework.

\subsection{Embedding Geometry and Isotropy}
\label{sec:rw_geometry}

The geometry of learned embedding spaces has long been studied as an indicator of representation quality. \citet{ethayarajh-2019-contextual} showed that contextual representations from pre-trained language models often occupy a narrow cone in high-dimensional space, a phenomenon known as \textit{anisotropy}, which can reduce discriminability. Subsequent work proposed several geometric diagnostics. In particular, prior studies have characterized isotropy using average pairwise cosine similarity and principal-component-based analyses~\cite{GaoHTQWL19,CaiHB021_Isotropy}, while \citet{RudmanGRE22_IsoScore,RudmanE24_stable_anisotropic} introduced IsoScore to quantify the uniformity of embedding space utilization. Although these properties have often been associated with representation quality~\cite{xiao-etal-2023-isotropy,JungPCKR23,godey-etal-2024-anisotropy}, more recent work suggests that their relationship with downstream performance is complex and not yet causally established~\cite{fuster-baggetto-fresno-2022-anisotropy,ait-saada-nadif-2023-anisotropy,RudmanE24_stable_anisotropic}.

In dense retrieval, anisotropy may reduce the effective capacity of the embedding space and thereby increase vulnerability to perturbations. Yet the relationship between embedding geometry and retrieval robustness remains underexplored, especially for modern LLM-based retrievers. It is therefore unclear whether properties such as angular uniformity and variance uniformity serve as reliable indicators of robustness, and which facets of robustness they may explain.\looseness=-1

\section{Methodology}

This section introduces the shared dense retrieval formulation used throughout the paper and the statistical methodology used specifically for the generalizability analysis in RQ1. RQ2 and RQ3 rely on the same retrieval pipeline, but their robustness-specific settings are introduced in the corresponding sections. Because evaluation datasets differ substantially in query volume, corpus size, and inherent difficulty, simple averaging conflates model capability with dataset-specific characteristics; we address this through linear mixed-effects models, which disentangle these confounds and enable more principled cross-dataset comparisons.\looseness=-1

\subsection{Dense Retrieval Formulation}
\label{sec:lm_dense_retrieval}

\paragraph{Formulation.}
Given a query $q$ and a document $d$ in a corpus $D$, a neural bi-encoder $E(\cdot)$ maps inputs into a shared $\hbar$-dimensional latent space, yielding embeddings $e_q = \text{Pooling}(E(q)) \in \mathbb{R}^{\hbar}$ and $e_d = \text{Pooling}(E(d)) \in \mathbb{R}^{\hbar}$. The pooling strategy depends on the architecture: BERT-based models typically use \texttt{[CLS]} or mean pooling, while decoder-only LLMs typically rely on \textbf{last-token pooling} or mean pooling due to their causal attention mechanism. In this study, we use each model's default pooling configuration as specified by its designers; pooling strategy is not an experimental variable. For instruction-augmented LLM-based retrievers, queries are prefixed with a task description before encoding to improve cross-task generalization. Relevance is scored via cosine similarity or dot product between $e_q$ and $e_d$. These document embeddings also form the basis of our geometric analyses in RQ3.

\paragraph{Inference and Evaluation Context.}
Dense retrievers are typically trained with contrastive learning~\cite{KarpukhinOMLWEC20_DPR}, often enhanced by knowledge distillation~\cite{HofstatterLYLH21_tasb_dense_retrieval}. We note this here only as background: in our experiments, we use official pretrained checkpoints rather than training retrievers from scratch. At inference time, document embeddings are pre-computed and indexed for retrieval. In large-scale deployments, this step is often implemented with Approximate Nearest Neighbor~(ANN) search~\cite{douze2024faiss}; in our experiments, however, the benchmark collections are small enough to permit exact search in FAISS. Shared evaluation metrics and implementation details are described in Section~4, while RQ-specific robustness metrics are introduced in the corresponding sections.

\subsection{Linear Mixed-Effects Analysis for Generalizability}
\label{sec:lmm_method}

With the shared retrieval pipeline established, we now introduce the statistical framework used to compare model generalizability across heterogeneous datasets. Although Analysis of Variance~(ANOVA) is commonly used to decompose the effects of multiple factors in IR evaluation~\cite{RobertsonK12,FerroS17,Ferro17,Ferro_0001S22,CulpepperFFK22}, our setting departs from ANOVA's assumptions due to two properties of the data: (1)~the evaluation is highly unbalanced, with different numbers of queries across datasets, and (2)~queries are nested within datasets, which induces statistical dependence.

To address these issues, we employ linear mixed-effects models~(LMMs)~\cite{Gałecki2013_LMM}, which extend classical linear models by explicitly modeling random variability due to hierarchical structure. For the factor-wise generalizability analyses in RQ1, we fit three LMMs---one for each of the three factor classifications: Task Type, Query Type, and Corpus Source Type. All three share the following form:
\begin{equation}
\label{eq:lmm_general}
\begin{split}
    y_{ikl} &= \underbrace{\beta_0 + \beta^{\textsc{m}}_i + \beta^{\textsc{f}}_{j}
    + \beta^{\textsc{m}\times\textsc{f}}_{ij}}_{\text{Fixed Effects}} \\
    &\quad + \underbrace{u_k + v_{kl}}_{\text{Random Effects}}
    + \underbrace{\epsilon_{ikl}}_{\text{Residual Error}},
\end{split}
\end{equation}
where $y_{ikl}$ is the query-level retrieval effectiveness score for model $i$ on query $l$ in dataset $k$ (nDCG@10 in this study), $\beta_0$ is the global intercept, and $j$ indexes the factor level associated with that observation. The fixed effects capture the systematic patterns of interest: $\beta^{\textsc{m}}_i$ denotes the main effect of model identity, $\beta^{\textsc{f}}_{j}$ denotes the main effect of the factor level $j$, and $\beta^{\textsc{m}\times\textsc{f}}_{ij}$ captures their interaction.

The interpretation of $j$ depends on the analysis. For Query Type, $j$ is defined at the query level; for Task Type and Corpus Source Type, it is determined by the dataset to which the query belongs. The random effects model the hierarchical structure of the data: $u_k \sim \mathcal{N}(0,\sigma_u^2)$ represents the random intercept for dataset $k$, capturing dataset-specific difficulty. We additionally include a query-level random intercept $v_{kl} \sim \mathcal{N}(0,\sigma_v^2)$ for query $l$ nested within dataset $k$; because the same query is evaluated for every model, this term is shared across models and captures query-specific difficulty beyond the dataset average. Finally, $\epsilon_{ikl} \sim \mathcal{N}(0,\sigma^2)$ is the residual error term.

From the fitted models, we derive \textbf{Estimated Marginal Means~(EMMs)} for each model-factor combination. These are population-level expected performance estimates obtained by evaluating the fixed-effects component of the model while averaging over the random effects. Intuitively, an EMM answers the question: ``What performance would this retriever achieve under a given condition, on a dataset of average difficulty?'' This provides a more standardized basis for comparing generalizability than raw averages alone. Statistical significance of fixed effects is assessed using Type III F-tests with Satterthwaite's approximation for degrees of freedom.

\section{Experimental Setup}

This section summarizes the experimental components shared across the paper. RQ-specific settings, such as the dataset taxonomy used in RQ1, the perturbation and attack configurations for RQ2, and the isotropy measurement protocols for RQ3, are introduced in the corresponding sections.

\subsection{Datasets}


\begin{table*}[t] 
\small
\centering
\caption{Statistics of datasets used in our work. Domain KB stands for Domain-Specific Knowledge Base. Avg. D/Q indicates the average number of relevant documents per query. Note that we report statistics exclusively for the evaluation test splits, ignoring training sets even where available.} 
\label{tab:datasets_details}
\renewcommand\arraystretch{1.1} %
\setlength{\tabcolsep}{2.5pt}{ %
 
\begin{tabular}{lllrrrrrcc }
\toprule
 \multirow{2}{*}{\textbf{Datasets}}  & \multirow{2}{*}{\textbf{Task}} & \multirow{2}{*}{\textbf{Corpus Source}}    & \multirow{2}{*}{\textbf{\#Corpus}}      &\multicolumn{2}{c}{\textbf{Test or Dev}} &\multicolumn{2}{c}{\textbf{Avg. Word Lengths}} 
    \\\cmidrule(lr){5-6}\cmidrule(lr){7-8} 
           &      &  & &\textbf{\#Query} &\textbf{Avg.D/Q}   &\textbf{Query}  &\textbf{Document} \\ 
\hline
\multicolumn{8}{c}{\textbf{MS MARCO}~(License: Custom Research License (Non-Commercial)) } \\
\hline
MS MARCO   &Passage Retrieval & General Web    &8,841,823  &6,980  &1.1 &5.96&55.98 \\
TREC DL 19   &Passage Retrieval &General Web   &8,841,823   &43  &95.4 &5.96&55.98 \\
TREC DL 20   &Passage Retrieval &General Web   &8,841,823   &54  &66.8 &5.96 &55.98 \\
\hline
 
\multicolumn{8}{c}{\textbf{BEIR  Benchmark}~(License: Various Research Licenses) } \\
\hline

TREC-COVID & Bio-Medical Retrieval  & Scientific Paper    &171,332  & 50&493.5 &10.6 &160.8  \\
NFCorpus & Bio-Medical Retrieval &Scientific Paper       &3,633   &323 &38.2 &3.3 &232.3 \\
NQ   &Question Answering  & Wikipedia      &2,681,468   &3,452 &1.2 &9.2 &78.9   \\ 
HotpotQA &Question Answering &Wikipedia       &5,233,329  &7,405 &2.0 & 17.6& 46.3\\ 
FiQA & Question Answering &Online Community      &57,638  &648 &2.6 &10.8 &132.3 \\ 
Quora   & Duplicate Question Retrieval  &Online Community      &522,931   & 10,000&1.6 &9.5 &11.4 \\
DBPedia & Entity Retrieval & Wikipedia      &4,635,922  & 400 &38.2 &5.4 &49.7 \\
SCIDOCS & Citation Prediction & Scientific Paper        &25,657   &1,000 &4.9 &9.4 &176.2 \\
SciFact & Fact Checking & Scientific Paper      &5,183   &300 &1.1 &12.4 &213.6   \\
FEVER  & Fact Checking &Wikipedia        &5,416,568   &6,666 &1.2 &8.1 &84.8  \\
Climate-FEVER & Fact Checking &Wikipedia      &5,416,593  &1,535 &3.0 &20.1 &84.8 \\
ArguAna & Argument Retrieval &Domain KB       &8,674   &1,406 &1.0 &193.0 &166.8  \\
Touché-2020 & Argument Retrieval &Domain KB        &382,545  &49 &19.0 &6.6 &292.4 \\ 
\hline
\multicolumn{8}{c}{\textbf{Bright  Benchmark}~(License: CC-BY-4.0) } \\
\hline
Biology & StackExchange Post Retrieval &Online Community      & 57,359 &103  &3.6  &115.2  & 83.6     \\
Earth Science &StackExchange Post Retrieval  &Online Community      &121,249  &116  &5.3  &109.5    &132.6      \\
Economics &StackExchange Post Retrieval  &Online Community       &50,220  &103  &8.0  &181.5  &120.2          \\
 Psychology &StackExchange Post Retrieval &Online Community        &52,835  &101  &7.3  &149.6  &118.2         \\
Robotics & StackExchange Post Retrieval  & Online Community      &61,961  &101  &5.5  &818.9  & 121.0      \\
Stack Overflow &StackExchange  Post Retrieval   &Online Community       & 107,081 &117  &7.0  &478.3  &704.7      \\
 Sustainable Living &StackExchange Post Retrieval   &Online Community      & 60,792 &108  & 5.6 &148.5  &107.9         \\
LeetCode     & Code Retrieval &Domain KB      & 413,932 &142  &1.8  &497.5&482.6       \\
Pony    & Code Retrieval &Domain KB       &7,894  &112  &22.5  & 102.6 &98.3          \\
AoPS     &Theorem Retrieval &Domain KB      &188,002  &111  &4.7  &117.1 & 250.5      \\
TheoremQA-Q  &Theorem Retrieval  & Domain KB       & 188,002 &194  &3.2  & 93.4 & 250.5        \\
TheoremQA-T    &Theorem Retrieval &Domain KB         &23,839  &76  &2.0  & 91.7 & 354.8    \\
\hline
\multicolumn{8}{c}{\textbf{BrowseComp-Plus Benchmark }~(License: CC-BY 4.0)}   \\
\hline
 Evidence & Question Answering  &General Web      &100,195  &830  & 6.1 &96.5  & 5179.2       \\
Gold & Question Answering   & General Web   &100,195  &830  & 2.9 &96.5  & 5179.2 \\
\bottomrule
\end{tabular} 
}
\end{table*}

We draw our evaluation datasets from four benchmarks: \textbf{MS MARCO}~\cite{bajaj2016ms} for in-domain passage retrieval, \textbf{BEIR}~\cite{thakur2021beir} for heterogeneous zero-shot transfer, \textbf{BRIGHT}~\cite{su2024bright} for reasoning-intensive retrieval, and \textbf{BrowseComp-Plus}~\cite{chen2025browsecomp} for web-scale retrieval with long and noisy documents. Together, these benchmarks cover 30 datasets spanning a broad range of tasks, domains, and corpus sources. Each RQ uses the relevant subset of these datasets, as described in the corresponding section. Table~\ref{tab:datasets_details} reports the full dataset statistics.

\subsection{Retrieval Models}

\begin{table*}[t]
\centering
\scriptsize
\caption{Overview of the retrieval models evaluated in this study.}
\label{tab:model_overview}
\renewcommand\arraystretch{1.2}
\setlength{\tabcolsep}{5.2pt}{
\begin{tabular}{l|rrrcllcc}
\thickhline
\multirow{2}{*}{\makecell[cc]{\textbf{Model} \\ \textbf{Name}}}  & \multirow{2}{*}{\makecell[cc]{\textbf{Model} \\ \textbf{Size}}} & \multirow{2}{*}{\makecell[cc]{\textbf{Embedding} \\ \textbf{Dimension}}} & \multirow{2}{*}{\makecell[cc]{\textbf{Context} \\ \textbf{ Length}}} &\multirow{2}{*}{\textbf{Huggingface Name}}  &\multirow{2}{*}{\makecell[cc]{\textbf{Release} \\ \textbf{Date}}}   &\multirow{2}{*}{\makecell[cc]{\textbf{Score} \\ \textbf{Function}}} &\multirow{2}{*}{\textbf{Pooling Strategy}} & \multirow{2}{*}{\textbf{License}}  \\  \\\hline
BM25  & N/A & N/A &N/A  & N/A  & N/A  & N/A & N/A& N/A\\ 
\hline
\multicolumn{7}{c}{\textit{Instruction-free Embedding Models}} \\
\hline
Contriever   & 110M & 768 &512 &\href{https://huggingface.co/facebook/contriever-msmarco}{contriever-msmarco}  & 2022 Jan & Inner Product & Mean Pooling & CC BY-NC 4.0 \\ 
BGE-M3   &561M  &1024  &8192 &\href{https://huggingface.co/BAAI/bge-m3}{bge-m3}& 2024 Jan & Cosine Similarity & CLS Pooling & MIT \\ 
\hline
\multicolumn{7}{c}{\textit{LLM-based Instruction-augmented Embedding Models}} \\ \hline
Qwen3   & 8B &4096 &32k & \href{https://huggingface.co/Qwen/Qwen3-Embedding-8B}{Qwen3-Embedding-8B} & 2025 Jul & Cosine Similarity &Last Token Pooling & Apache-2.0\\ 
Linq   & 7B &4096 &32k  & \href{https://huggingface.co/Linq-AI-Research/Linq-Embed-Mistral}{Linq-Embed-Mistral} & 2024 May & Cosine Similarity & Last Token Pooling& CC BY-NC 4.0\\ 
GTE   & 7B &3584 &32k & \href{https://huggingface.co/Alibaba-NLP/gte-Qwen2-7B-instruct}{gte-Qwen2-7B-instruct} & 2024 Jun & Cosine Similarity & Last Token Pooling & Apache-2.0\\ 
\hline
\multicolumn{7}{c}{\textit{LLM-based Instruction-augmented Reasoning Embedding Models}} \\
\hline
ReasonIR   & 8B &4096 &128k  &\href{https://huggingface.co/reasonir/ReasonIR-8B}{ReasonIR-8B} & 2025 Apr & Cosine Similarity & Mean Pooling & CC BY-NC 4.0\\ 
DIVER   & 4B  &2560 &40k & \href{https://huggingface.co/AQ-MedAI/Diver-Retriever-4B}{Diver-Retriever-4B} & 2025 Aug & Cosine Similarity & Last Token Pooling & Apache-2.0\\ 
ReasonEmbed& 8B &4096 &32k & 
\href{https://huggingface.co/hanhainebula/reason-embed-qwen3-8b-0928}{reason-embed-qwen3-8b-0928} &2025 Oct & Cosine Similarity & Last Token Pooling & Apache-2.0 \\
\hline

\end{tabular}}
\end{table*}

We evaluate a shared pool of retrievers spanning three categories. First, we include the classical lexical baseline \textbf{BM25}~\cite{RobertsonZ09_BM25}. Second, we include two encoder-only dense baselines, \textbf{Contriever}~\cite{izacard2021contriever} and \textbf{BGE-M3}~\cite{chen-etal-2024-m3}. Third, we evaluate six decoder-only LLM-based retrievers: three general-purpose instruction-augmented models, \textbf{Qwen3}~\cite{qwen3embedding}, \textbf{Linq}~\cite{choi2024linq}, and \textbf{GTE}~\cite{li2023towards}, and three reasoning-oriented models, \textbf{ReasonIR}~\cite{shao2025reasonir}, \textbf{DIVER}~\cite{long2025diver}, and \textbf{ReasonEmbed}~\cite{chen2025reasonembedenhancedtextembeddings}. Their main architectural and implementation characteristics are summarized in Table~\ref{tab:model_overview}. These nine models constitute the shared core pool used in the main cross-model comparisons; any RQ-specific exclusions or auxiliary additions are noted in the corresponding sections.

\subsection{Implementation Details and Evaluation Metrics}

Unless otherwise noted, we use the official checkpoints released by the model developers and follow each model's default encoding configuration, including its recommended pooling strategy and prompt format when applicable. For instruction-augmented retrievers, we use the developers' default query prompt templates; the exact templates and preprocessing settings are documented in our code repository. To standardize long-document handling, we truncate document inputs to 512 tokens for Contriever and to 8192 tokens for all other dense retrievers. All dense retrievers use cosine similarity as the relevance scoring function, with the exception of Contriever, which uses dot product following its original design. In our experiments, document embeddings are pre-computed and searched exactly using FAISS~\cite{douze2024faiss}; the remaining pipeline follows the standard dense retrieval pipeline described in Section~\ref{sec:lm_dense_retrieval}. For BM25, we use the \texttt{LuceneBM25Model} implementation from Gensim with $k_1=0.9$ and $b=0.4$, consistent with the setup used in our shared evaluation pipeline.

We use \textbf{nDCG@10} as the primary retrieval effectiveness metric throughout the paper. Robustness-specific metrics are introduced in the relevant sections: RQ2 reports relative performance drop under query perturbations and ASR@20 under corpus poisoning, whereas RQ3 analyzes the relationship between robustness outcomes and model-side properties derived from the same shared retrieval pipeline. All experiments were conducted on Snellius, the Dutch National Supercomputer, using NVIDIA H100 GPUs. Full implementation details and environment configurations are provided 
in our code repository.\footnote{\url{https://github.com/liyongkang123/Robust_LLM_Retriever_Eval}}

\section{RQ1: How generalizable are LLM-based dense retrievers across diverse conditions?}
 
We address RQ1 through a structured evaluation of retriever generalizability across diverse tasks, query styles, and corpus sources.

\subsection{RQ1-Specific Setup}

RQ1 uses all nine retrievers and all 30 datasets described in Section~4. The analysis proceeds in two stages. We first report raw per-dataset nDCG@10 scores to establish a baseline picture of cross-dataset performance heterogeneity. We then apply the linear mixed-effects methodology described in Section~\ref{sec:lmm_method} to derive Estimated Marginal Means~(EMMs) that control for dataset- and query-level confounds. To support this factor-conditional analysis, we define a three-factor taxonomy of the evaluation space below.

\subsubsection{Taxonomy}
We organize the evaluation space along three factors: \textit{task type}, \textit{query type}, and \textit{corpus source type}. The first two are standard abstractions in retrieval evaluation, while the third captures variation in corpus provenance that may affect transfer. Dataset-level mappings for task type and corpus source type are summarized in Table~\ref{tab:datasets_details}, while query types are assigned automatically using the classifier described below.

\begin{table*}[t]
\small
\centering
\caption{The query type taxonomy we used in this paper, which comes from~\citep{BolotovaBSCS22_query_taxonomy}. The percentage below each category name indicates the proportion of queries assigned to that type, computed over all queries in the 30 datasets drawn from the four benchmarks used in this paper.}
\label{tab:nfq_taxonomy}
\begin{tabularx}{\textwidth}{>{\centering\arraybackslash}p{3.5cm}|X|X|X}
    \hline
    \textbf{Category} & \textbf{Description} & \textbf{Expected Answer Structure} & \textbf{Patterns} \\
    \hline
    \multirow{3}{*}{\makecell[c]{\textbf{FACTOID} \\ \small{(37.4\%)}}} & You want to find a \textbf{specific, factual piece of information}, such as a date, a location, or a name. & A concise factual statement or direct answer, often one or a few words. & Where is ... located? \newline Who is/was ...?\newline When did/was ...?\\
    \hline
    \multirow{4}{*}{\makecell[c]{\textbf{INSTRUCTION} \\ \small{(6.4\%)}}} & You want to understand the \textbf{procedure/method} of doing/achieving something. & Instructions/guidelines provided in a step-by-step manner. & How to ...? \newline How can I do ...? \newline What is the process for ...? \newline What is the best way to ...? \\
    \hline
    \multirow{4}{*}{\makecell[c]{\textbf{REASON} \\ \small{(3.2\%)}}} & You want to find out \textbf{reasons} of/for something. & A list of reasons with evidence. & Why does ...? \newline What is the reason for ...? \newline What causes ...? \newline How come ... happened? \\
    \hline
    \multirow{5}{*}{\makecell[c]{\textbf{EVIDENCE-BASED} \\ \small{(13.1\%)}}} & You want to learn about the \textbf{features/description/definition} of a concept/idea/object/event. & Wikipedia-like passage describing/defining an event/object or its properties based only on facts. & What is ...? \newline How does/do ... work? \newline What are the properties of ...? \newline What is the meaning of ...? \newline How do you describe ...? \\
    \hline
    \multirow{3}{*}{\makecell[c]{\textbf{COMPARISON} \\ \small{(1.3\%)}}} & You want to \textbf{compare/contrast} two or more things, understand their differences/similarities. & A list of key differences and/or similarities of something compared to another thing. & How is X ... to/from Y? \newline What are the ... of X over Y? \newline How does X ... against Y? \\
    \hline
    \multirow{5}{*}{\makecell[c]{\textbf{EXPERIENCE} \\ \small{(3.7\%)}}} & You want to get \textbf{advice} or \textbf{recommendations} on a particular topic. & Advantages, disadvantages, and main features of an entity (product, event, person, etc) summarised from personal experiences. & Would you recommend ...? \newline How do you like ...? \newline What do you think about ...? \newline Should I ...? \\
    \hline
    \multirow{5}{*}{\makecell[c]{\textbf{DEBATE} \\ \small{(7.7\%)}}} & You want to \textbf{debate on a hypothetical question} (is someone right or wrong, is some event perceived positively or negatively?). & Arguments on a debatable topic consisting of different opinions on something supported or weakened by pros and cons of the topic in the question. & Does ... exist? \newline Can ... be successful? \newline Do you think ... are ...? \newline Is ... really a ...? \\
    \hline
    \multirow{6}{*}{\makecell[c]{\textbf{NOT-A-QUESTION} \\ \small{(27.2\%)}}} & These queries \textbf{do not explicitly ask a question} but instead may express a need for information, a request for clarification, or a search for specific content without the use of an interrogative form. & The response might include a document, a summary, a list, or relevant content that aligns with the user's implied request or interest. The answer is often less direct than in other query types. & Find ... in ... \newline Details on ...\newline Gmail login \newline Specific brand, person, etc. \\
    \hline
\end{tabularx}
\end{table*}

\paragraph{Factor 1: Task Types.} 
We adopt the BEIR task definitions~\cite{thakur2021beir} as the starting point for shared labels across benchmarks. For datasets outside BEIR, we classify BrowseComp-Plus as Question Answering and group the 12 BRIGHT datasets into StackExchange Post Retrieval, Code Retrieval, and Theorem Retrieval following their original benchmark design~\cite{su2024bright}. This yields 11 task types in total: Passage Retrieval, Bio-Medical Retrieval, Question Answering, Duplicate Question Retrieval, Entity Retrieval, Citation Prediction, Fact Checking, Argument Retrieval, StackExchange Post Retrieval, Code Retrieval, and Theorem Retrieval.

\paragraph{Factor 2: Query Types.} 
We use the query-intent taxonomy of \citet{BolotovaBSCS22_query_taxonomy} and apply their open-source classifier\footnote{\url{https://huggingface.co/Lurunchik/nf-cats}} to assign each query to one of eight types: FACTOID, INSTRUCTION, REASON, EVIDENCE-BASED, COMPARISON, EXPERIENCE, DEBATE, and NOT-A-QUESTION. Table~\ref{tab:nfq_taxonomy} summarizes the category definitions and their proportions across our 30 datasets. The distribution is heavily skewed toward FACTOID and NOT-A-QUESTION queries, which together account for nearly two-thirds of the full evaluation suite.


\paragraph{Factor 3: Corpus Source Types.} 
By examining the data construction process of each dataset, we classify corpora into five source categories: General Web, Wikipedia, Scientific Paper, Online Community, and Domain KB. General Web covers open-web crawls with heterogeneous noise and topic distributions; Wikipedia captures relatively clean encyclopedic collections; Scientific Paper covers academic articles and abstracts; Online Community contains user-generated forum content such as Reddit and StackExchange; and Domain KB refers to curated, domain-specific knowledge bases such as debate portals, coding platforms, and theorem repositories.



\subsection{Raw Per-Dataset Results}



\begin{table*}[t]
\centering
\small
\caption{Raw nDCG@10 scores for all models across all benchmarks. Bold indicates best per row. Note that the linear mixed-effects models in the main text use query-level scores rather than these aggregated results.}
\label{tab:main_experiments_results}
\renewcommand\arraystretch{1} %
\setlength{\tabcolsep}{1mm}{ %
\begin{tabular}{llcccccccccc}
\hline
  \multirow{2}{*}{\textbf{Dataset}} & & \multicolumn{2}{c}{\textbf{Instruction-free}}& \multicolumn{6}{c}{\textbf{Instruction-augmented}} \\ \cmidrule(lr){3-4} \cmidrule(lr){5-10} & \textbf{BM25}   & \textbf{Contriever}  &\textbf{BGE-M3} & \textbf{Qwen3} & \textbf{Linq} & \textbf{GTE} & \textbf{ReasonIR} &\textbf{DIVER} &  \textbf{ReasonEmbed} \\
\hline
\multicolumn{10}{l}{\textit{MS MARCO}} \\ 
MS MARCO  & 21.8   &40.7    &38.4  &43.7   &44.9   &46.0    &32.2   & 31.4   &32.3   \\
TREC DL 19 & 50.0   &67.6   &67.3  & 76.3 & 74.6 & 74.4  &63.9  &67.7  &68.8   \\
TREC DL 20 & 48.7   &66.7   &67.8  &74.3   & 73.0  & 72.9  &63.8  & 64.4 &64.6   \\
\hdashline
\multicolumn{1}{c}{\textbf{Avg. (MS MARCO)}}  & 40.2 & 58.3 & 57.8 & 64.8 & 64.2 & 64.4 & 53.3 & 54.5 & 55.2 \\
\hline
\multicolumn{10}{l}{\textit{BEIR}} \\ 
TREC-COVID   & 52.9   & {59.4} &55.6 &\textbf{94.7} &87.1 &81.1 &76.8 &78.8 & 81.2       \\
NFCorpus  &31.4    &32.8 &31.6 & 41.4  &\textbf{41.9} &39.9 &37.9 &38.0 &  39.9      \\
NQ & 29.5  &49.8 & 60.6 &64.9  &\textbf{70.4} &66.8 &52.6 & 54.2  &60.7      \\
HotpotQA      & 60.1   &63.8 &69.5 &\textbf{76.8} &76.4 &73.0 &62.8 &65.8 & 44.4       \\
FiQA     & 23.7    &33.0 &41.1 &\textbf{64.7} &61.2 &61.8   &48.4 &47.9 &44.2        \\
Quora        & 76.9   &86.4 &88.6 &89.0 &\textbf{90.2} &89.6 &81.0 &88.4 & 48.5       \\
DBPedia     & 31.0  & 41.3 & 39.8 & 49.9  &51.3  &52.4  &39.7  &41.5 & 41.3       \\
SCIDOCS      & 14.4  &16.6 & 16.4 &\textbf{32.8} &22.1 &28.9 &22.8 &23.5 & 22.5  \\
SciFact   &  67.0    &67.7 & 64.1 &78.5 &78.7 &\textbf{79.5} &75.1 &76.6 & 78.8      \\
FEVER   & 60.3    &75.7 &80.9 &91.8 &92.3 &\textbf{95.2} &77.4 & 77.8 &  65.1      \\
Climate-FEVER    & 15.2   &23.8 &29.7 &47.4 &39.3 &46.2 &31.4 &30.2 & 25.2       \\
ArguAna    & 38.7  & 44.5  &54.0 &\textbf{76.6} &68.2 &68.5 &60.2 &48.3 & 49.0       \\
Touché-2020     & \textbf{44.4}        &20.5 &22.1 &35.7 &30.0 &33.8 &19.6 &24.8 & 26.5  \\
\hdashline
\multicolumn{1}{c}{\textbf{Avg. (BEIR)}}  &42.0 &47.3 &50.3  &\textbf{64.9}  &62.2  &62.8  &52.7  &53.5 &48.3   \\
\hline
\multicolumn{10}{l}{\textit{BRIGHT}} \\ 
Biology &18.9   &19.0 &10.3 &19.7 &25.2 &30.3 &26.2 & 41.9  &\textbf{53.2}       \\
Earth Science &27.2  &28.3 &17.5 &32.5 &31.4 &40.3 &31.1 & 44.6  &\textbf{56.4}        \\
Economics & 14.9    & 17.4 &11.1 &18.2 &22.9 &15.8 & 23.5  &22.0 &\textbf{35.5}        \\
Psychology & 12.5    &16.2 &14.5 &27.4 &26.7 &26.4 &30.1 & 34.4  &\textbf{46.2}       \\
Robotics & 13.6    &14.1 &11.2 &15.9 &18.3 &12.5 &18.2 &20.9 & \textbf{35.8}       \\
Stack Overflow &   18.4  &14.2 &9.1 &18.4 &16.6 &15.8 & 24.0  &20.8 & \textbf{36.7}    \\
 Sustainable Living & 15.0   &14.3 &10.0 &17.5 &24.8 &20.9 &20.1 &25.3 &  \textbf{39.2}     \\
LeetCode  & 24.1    &22.2 &24.7 &33.8 &30.9 &31.0 &34.2 &\textbf{37.7} & 34.2       \\
Pony &7.9     &9.8 &14.8 &1.0 &\textbf{23.7} &2.5 &9.9 &13.6 & 17.9       \\
AoPS  &6.2    &9.8 &4.6 &9.7 &8.2 &\textbf{15.8} &14.6 &11.1 &12.9        \\
TheoremQA-Q & 10.4    &10.1 &12.9 &39.0 &28.9 &32.2 &31.7 &38.5 &  \textbf{41.0}     \\
TheoremQA-T  & 4.9    &4.0 &5.2 &39.3 &33.0 &35.8 &27.8 & 36.8 &  \textbf{45.8}      \\
\hdashline
\multicolumn{1}{c}{\textbf{Avg. (Bright)}} &14.5 &15.0  &12.2  &22.7  &24.2  &23.3  &24.3  &29.0 & \textbf{37.9}  \\
\hline
\multicolumn{10}{l}{\textit{BrowseComp-Plus}} \\ 
 Evidence &4.4  &6.9 &6.3 &\textbf{20.1} &12.4 &17.4 &15.5 &17.7 &16.1        \\
 Gold &3.7     &6.5 &5.1 &\textbf{18.7} &10.8 &17.8 &13.6 &17.3 & 16.9       \\
\hdashline
\multicolumn{1}{c}{\textbf{Avg. (BrowseComp-Plus)}}  &4.1 &6.7  &5.7  &19.4  &11.6  &17.6  &14.6  &17.5 &16.5   \\
\hline
\end{tabular}}
\end{table*}

Table~\ref{tab:main_experiments_results} reports the per-dataset nDCG@10 scores that underlie all aggregated analyses in the main text. The raw results exhibit substantial heterogeneity in retrieval difficulty, with scores ranging from above 90 on comparatively easy, high-signal datasets (e.g., FEVER, Quora) to near zero on the most challenging settings (e.g., BrowseComp-Plus, where even the best model scores below 5). This variability highlights that naive aggregation of raw scores can be strongly confounded by dataset-specific characteristics, motivating the difficulty-adjusted comparisons provided by our linear mixed-effects analysis.\looseness=-1

Several patterns emerge from the raw scores. First, performance varies systematically across benchmarks: models achieve their highest scores on BEIR, followed by MS~MARCO, with BRIGHT and BrowseComp-Plus proving most challenging. The average nDCG@10 gap between BEIR and BRIGHT exceeds 30 points for most models, reflecting the qualitatively different retrieval demands of reasoning-intensive queries. Second, instruction-augmented LLM retrievers (Qwen3, Linq, and GTE) generally lead across the evaluation suite, while reasoning-oriented models exhibit a pronounced specialization pattern: ReasonEmbed achieves the best performance on 9 of 12 BRIGHT datasets but underperforms sharply on standard datasets such as Quora (48.5 vs.\ $>$88 for other LLM retrievers). Third, even SOTA dense retrievers struggle on specialized domains such as code and theorem retrieval, where nDCG@10 scores remain uniformly low across all models (typically below 30).

Lexical baselines also remain non-trivial competitors in specific regimes. BM25 performs strongly on argument-centric retrieval, notably Touch\'e-2020, and remains comparable to encoder-only dense models on BRIGHT on average (14.5 vs.\ 15.0 for Contriever and 12.2 for BGE-M3). These observations suggest that exact-match signals still matter for some reasoning-heavy or domain-curated corpora. At the same time, the large variation across datasets makes raw scores alone insufficient for fair factor-level comparison, so we next turn to the mixed-effects analysis introduced in Section~\ref{sec:lmm_method}.

\subsection{Linear Mixed-Effects Results}

We now apply the linear mixed-effects methodology described in Section~\ref{sec:lmm_method} to quantify retriever generalizability while controlling for dataset- and query-level heterogeneity. We first examine joint tests of fixed effects to determine whether model differences and model-factor interactions are statistically reliable, and then interpret the resulting Estimated Marginal Means~(EMMs) for each factor level.

\begin{table}[!t]
    \centering
    \small
    \caption{Joint (Type III) F-tests of fixed effects with Satterthwaite approximation~\cite{JSSv082i13,lenth2023emmeans} for the three linear mixed-effects models. 
Significant interaction terms indicate non-uniform generalizability 
across factor levels.
}
    \label{tab:anova_joint_tests}
    \renewcommand\arraystretch{1}
    \setlength{\tabcolsep}{3.5pt} 
    \begin{tabular}{ll r r c}
    \toprule
   \textbf{Factor}& \textbf{Fixed Effects} & \textbf{$F$-ratio}  & \textbf{$p$-value} \\
    \midrule
    \multirow{3}{*}{\shortstack[l]{Task \\Type }}
   & Model & 675.89   & $< .001$ \\
    &Task Type & 2.67   & \phantom{$<$}.003 \\
    &Model $\times$ Task Type & 303.48   & $< .001$ \\
    \midrule
    \multirow{3}{*}{\shortstack[l]{Query \\Type }}
    &Model & 3143.25   & $< .001$ \\
    &Query Type & 32.85   & $< .001$ \\
   & Model $\times$ Query Type & 166.54   & $< .001$ \\
    \midrule
    \multirow{3}{*}{\shortstack[l]{Corpus\\Source\\Type}}
    &Model & 2276.43   & $< .001$ \\
    &Corpus Source Type & 2.33   & \phantom{$<$}.054 \\
    &Model $\times$ Corpus Source Type & 468.79  & $< .001$ \\
    \bottomrule
    \end{tabular}
\end{table}

\paragraph{Statistical Significance of Factor Effects.}
Table~\ref{tab:anova_joint_tests} reports Type III F-tests of fixed effects for three linear mixed-effects model analyses. All Model $\times$ Factor interaction terms are highly significant ($p < .001$), with large $F$-ratios (ranging from 166.54 to 468.79), indicating that performance gaps between retrievers vary substantially across task types, query types, and corpus source types. A noteworthy asymmetry emerges for corpus source: while its main effect does not reach conventional significance ($F = 2.33$, $p = .054$), its interaction with model is the largest of the three analyses ($F = 468.79$). We interpret this pattern cautiously: corpus source does not induce a uniform average shift across retrievers, but the effect of corpus provenance differs sharply by model. This motivates examining corpus-source-specific EMMs rather than relying on a single pooled effect.

\begin{table*}[t]
\centering
\small
\caption{Estimated Marginal Means of nDCG@10 derived from linear mixed-effects models across task, query, and corpus source factors, controlling for dataset difficulty and scale. Results indicate that instruction-augmented retrievers achieve the strongest estimated generalizability across diverse factors. Bold indicates the best performance in each row.}
\label{tab:emm_merge_results}
\renewcommand\arraystretch{0.95}
\setlength{\tabcolsep}{1.5pt}
\begin{tabular}{llccccccccccccc}
\toprule
\multirow{2}{*}{\textbf{Factor}}&\multirow{2}{*}{\textbf{Level}} &
  & \multicolumn{2}{c}{\textbf{Instruction-free}} 
  & \multicolumn{6}{c}{\textbf{Instruction-augmented}} \\
\cmidrule(lr){4-5} \cmidrule(lr){6-11}
& & \textbf{BM25} & \textbf{Contriever} & \textbf{BGE-M3}
 & \textbf{Qwen3} & \textbf{Linq} & \textbf{GTE} & \textbf{ReasonIR} & \textbf{DIVER} & \textbf{ReasonEmbed}   \\
\midrule
 
\multirow{11}{*}{\shortstack[l]{Task \\Type }}
&Passage Retrieval  & 22.1 & 41.1 & 38.8 & 44.2 & 45.3 & \textbf{46.3} & 32.7 & 31.9 & 32.7 \\
&Bio-Medical Retrieval      & 47.5 & 49.6 & 48.1 & \textbf{61.7} & 61.2 & 58.7 & 56.4 & 56.8 & 58.7 \\
&Question Answering         & 32.8 & 41.4 & 48.2 & 56.0 & \textbf{56.6} & 53.9 & 43.0 & 45.3 & 34.0 \\
&Duplicate Question Retrieval & 76.9 & 86.4 & 88.6 & 89.0 & \textbf{90.2} & 89.6 & 81.0 & 88.4 & 48.5 \\
&Entity Retrieval    & 31.0 & 41.3 & 39.8 & 49.9 & 51.3 & \textbf{52.4} & 39.8 & 41.5 & 41.3 \\
&Citation Prediction  & 14.4 & 16.6 & 16.4 & \textbf{32.8} & 22.1 & 28.9 & 22.8 & 23.5 & 22.5 \\
&Fact Checking & 43.4 & 57.1 & 62.1 & 74.3 & 73.3 & \textbf{76.8} & 60.1 & 60.2 & 49.4 \\
&Argument Retrieval   & 26.1 & 31.0 & 40.2 & \textbf{62.5} & 54.1 & 54.6 & 46.1 & 34.8 & 35.5 \\
&StackExchange Post Retrieval   & 17.3 & 17.7 & 11.9 & 21.4 & {23.6} & 23.2 & 24.7 & 30.0 & \textbf{43.3} \\
&Code Retrieval & 15.8 & 15.6 & 19.2 & 18.2 & 26.6 & 17.3 & 22.4 & \textbf{26.0} & 25.9 \\
&Theorem Retrieval   & 7.2 & 7.9 & 8.0 & 29.6 & 22.8 & 27.2 & 25.0 & 29.3 & \textbf{32.9} \\
\hdashline
&Avg. (task)               & 30.4 & 36.9 & 38.3 & \textbf{49.1} & 47.9 & 48.1 & 41.3 & 42.5 & 38.6 \\

\midrule
 
\multirow{9}{*}{\shortstack[l]{Query \\Type }}
&FACTOID                     & 25.7 & 35.8 & 39.6 & 46.5 & \textbf{47.1} & 45.7 & 34.1 & 36.3 & 27.5 \\
&INSTRUCTION                 & 26.2 & 35.8 & 38.3 & 40.8 & \textbf{41.8} & \textbf{41.8} & 30.8 & 38.8 & 0.2 \\
&REASON                      & 30.2 & 39.6 & 41.6 & 44.4 & \textbf{46.1} & 44.9 & 39.1 & 43.0 & 16.4 \\
&EVIDENCE-BASED              & 23.4 & 39.5 & 39.8 & 44.3 & 45.0 & \textbf{45.5} & 35.5 & 35.9 & 22.2 \\
&COMPARISON                  & 36.9 & 48.0 & 52.4 & 55.1 & \textbf{55.8} & 54.1 & 44.8 & 48.8 & 26.1 \\
&EXPERIENCE                  & 29.9 & 40.2 & 43.0 & 43.3 & \textbf{44.9} & 44.2 & 33.4 & 42.8 & 0.8 \\
&DEBATE                      & 31.7 & 40.6 & 43.4 & 47.2 & \textbf{47.7} & 46.8 & 37.7 & 43.1 & 16.9 \\
&NOT-A-QUESTION              & 21.8 & 32.7 & 37.3 & 50.6 & 48.4 & \textbf{51.1} & 37.9 & 37.0 & 29.3 \\
\hdashline
&Avg. (query)                & 28.2 & 39.0 & 41.9 & 46.5 & \textbf{47.1} & 46.8 & 36.7 & 40.7 & 17.4 \\
\midrule
\multirow{6}{*}{\shortstack[l]{Corpus\\Source\\Type}}
&General Web & 10.7 & 27.9 & 25.8 & 32.0 & 32.3 & \textbf{33.7} & 21.3 & 20.8 & 21.4 \\
&Wikipedia  & 40.2 & 51.4 & 57.7 & 66.6 & \textbf{67.0} & 66.6 & 52.8 & 54.2 & 42.5 \\
&Scientific Paper   & 44.1 & 46.0 & 44.9 & \textbf{60.3} & 53.8 & 57.5 & 52.5 & 53.2 & 53.5 \\
&Online Community   & 26.6 & 35.5 & 37.4 & 39.8 & \textbf{40.8} & 40.3 & 32.1 & 38.9 & 4.5 \\
&Domain KB  & 12.4 & 15.9 & 22.8 & \textbf{42.2} & 36.1 & 36.1 & 30.4 & 23.7 & 24.9 \\ 
\hdashline
&Avg. (corpus source)               & 26.8 & 35.4 & 37.7 & \textbf{48.2} & 46.0 & 46.8 & 37.8 & 38.2 & 29.4 \\
\bottomrule
\end{tabular}
\end{table*}

\paragraph{Factor-Wise Performance Patterns.}
Table~\ref{tab:emm_merge_results} reports the resulting EMMs. Across all three factors, instruction-augmented LLM retrievers (Qwen3, Linq, and GTE) achieve the strongest average EMMs (task: 47.9--49.1; query: 46.5--47.1; corpus source: 46.0--48.2), consistently outperforming both BM25 and encoder-only baselines. Their advantage is broad rather than confined to a single factor, indicating stronger estimated generalizability across the heterogeneous conditions covered by our evaluation.

\textit{Task Type.} Performance varies substantially across the 11 task types. All models perform well on Duplicate Question Retrieval (EMMs $>$80 for LLM retrievers), whereas Citation Prediction and Theorem Retrieval remain the most challenging (Qwen3: 32.8 and 29.6, respectively). Within the BRIGHT-derived task types, reasoning-oriented models are especially competitive on theorem and community-style technical retrieval: ReasonEmbed leads on Theorem Retrieval (32.9) and StackExchange Post Retrieval (43.3). By contrast, Code Retrieval is more mixed, with Linq (26.6), DIVER (26.0), and ReasonEmbed (25.9) performing similarly. Citation Prediction remains led by Qwen3 (32.8), while ReasonEmbed is only competitive (22.5). Overall, the advantage of reasoning-oriented models is task-specific rather than uniform across all reasoning-intensive settings.

\textit{Query Type.} Instruction-augmented retrievers lead across all eight query types, with their advantage most pronounced on NOT-A-QUESTION queries (Qwen3: 50.6, GTE: 51.1 vs.\ BM25: 21.8). The most striking pattern involves ReasonEmbed, which achieves near-zero EMMs on INSTRUCTION (0.2) and EXPERIENCE (0.8) queries. This near-complete failure—far below BM25 (26.2 and 29.9)—suggests a pronounced mismatch between ReasonEmbed and queries classified into these categories. One plausible explanation is that its training recipe and prompt behavior are less well aligned with imperative or advice-seeking formulations than with fact-seeking or reasoning-oriented ones, although the present analysis cannot determine the source of the mismatch directly. ReasonIR and DIVER do not exhibit this failure mode, indicating that it is model-specific rather than a general property of reasoning-oriented training.

\textit{Corpus Source Type.} Wikipedia-sourced corpora yield the highest EMMs across all models (Qwen3: 66.6, BM25: 40.2), likely reflecting the prevalence of Wikipedia-based training data. General Web corpora induce a collective performance drop across all models (best: GTE 33.7), likely attributable to noisy and heterogeneous document distributions. Domain KB corpora reveal the largest gap between model tiers: Qwen3 achieves 42.2 versus BM25's 12.4, suggesting that instruction-augmented models handle domain-specific curated content far better than lexical baselines. Online Community corpora expose ReasonEmbed's strongest weakness (EMM: 4.5), consistent with its near-zero scores on forum-style INSTRUCTION and EXPERIENCE queries.

\textit{Summary of Generalizability Analysis.} Instruction-augmented LLM retrievers exhibit the strongest overall estimated generalizability across all three factors. Reasoning-focused retrievers exhibit what we term a \textit{specialization tax}: they achieve strong performance on narrow task niches (e.g., theorem and StackExchange post retrieval) while degrading markedly on several other query types and corpus sources. Corpus source shift remains a key challenge even for SOTA models, and corpus provenance interacts strongly with model identity rather than producing a uniform effect across retrievers.

\section{RQ2: How stable are LLM-based dense retrievers under perturbations?}\label{Sec:RQ2_main}

Beyond generalizability across conditions, practical deployment requires stability under input perturbations. We focus on neural dense retrievers throughout this section, excluding BM25. This is partly because BM25 is inapplicable to gradient-based corpus poisoning (it has no learnable parameters), and partly to maintain a consistent model set across both query-side and document-side analyses. We evaluate stability from two complementary perspectives: \textit{query-side stability} under five types of natural query variations that simulate realistic user behavior, and \textit{document-side stability} under corpus poisoning attacks that test robustness against adversarial corpus manipulation. Together, these analyses characterize how robustness varies across perturbation types, attack threat models, and retrieval architectures.

\begin{table*}[t]
\centering
\small
\caption{Examples of query variations for each perturbation type from the NQ dataset.}
\label{tab:query_variation_examples}
\begin{tabular}{p{2.5cm} p{6cm} p{6cm}}
\toprule
\textbf{Variation} & \textbf{Original Query} & \textbf{Perturbed Query} \\
\midrule
Misspelling 
    & who is the leader of the ontario pc party 
    & who is the leader of the ontario pc paryt\\
\addlinespace
Reordering 
    & who plays the doc in back to the future 
    & who back the doc in plays to the future \\
\addlinespace
Synonymizing 
    & when did the apple iphone se come out 
    & when did the apple iphone se arriving out \\
\addlinespace
Paraphrasing 
    & why does king from tekken wear a mask
    & Why the King of Tekken Wears a Mask \\
\addlinespace
Naturalizing 
    & where does the donkey talk in the bible 
    & donkey talk bible \\
\bottomrule
\end{tabular}
\end{table*}

\subsection{Query-Side Stability: Query Variations}

\subsubsection{Experimental Setup}
Realistic user queries deviate from idealized forms in various ways, ranging from unintentional typing errors to deliberate reformulations. To systematically assess how LLM-based retrievers handle such variation, we adopt five perturbation types from \citet{PenhaCH22_query_variation}\footnote{\url{https://github.com/Guzpenha/query_variation_generators}}, each targeting a distinct aspect of query robustness. We evaluate on four datasets (NQ, MS~MARCO Dev, HotpotQA, and FiQA), chosen to cover distinct query styles and difficulty regimes while keeping perturbation generation and repeated evaluation computationally manageable, and report the relative performance drop rate: $\text{Drop Rate} = (s_{\text{orig}} - s_{\text{perturbed}}) / s_{\text{orig}} \times 100\%$, where $s$ denotes nDCG@10. When a perturbation type admits multiple generation methods, we randomly select one method per query; all experiments use 5 random seeds (1999, 2016, 2026, 5, 27). Table~\ref{tab:query_variation_examples} provides examples of each perturbation type.

$\bullet$ \textbf{Misspelling}: Introducing typographical errors into query terms. We adopt the more comprehensive misspelling method from \citet{zhuang2021dealing,CharacterBERT}\footnote{\url{https://github.com/ielab/CharacterBERT-DR/tree/main/data}}, which covers a broader range of character-level operation types than the original method in \citet{PenhaCH22_query_variation}; it considers only keywords with more than 3 characters and modifies exactly one keyword per query. For each selected keyword, we randomly apply one of the following operations: character insertion, character deletion, character substitution, neighbor character swap, or adjacent keyboard character swap.

$\bullet$ \textbf{Reordering}: Randomly swapping two words in the query while preserving the original vocabulary.

$\bullet$ \textbf{Synonymizing}: Replacing a non-stopword with a synonym, randomly selected from one of two methods: (1) the nearest neighbor in counter-fitted GloVe embedding space, which yields better synonyms than standard GloVe embeddings; or (2) the first synonym found in WordNet.

$\bullet$ \textbf{Paraphrasing}: Generating semantically equivalent reformulations using one of two randomly selected methods: (1) back-translation through a pivot language; or (2) a T5 model fine-tuned for paraphrase generation.

$\bullet$ \textbf{Naturalizing}: Removing all stopwords from the query to simulate keyword-style search behavior. Note that \citet{PenhaCH22_query_variation} also proposed a T5-based method for generating natural language questions~(trained on TREC topic title-description pairs); however, the fine-tuned model is no longer publicly available\footnote{\url{https://github.com/Guzpenha/query_variation_generators/issues/3}}, so we use only the stopword removal approach.

\begin{figure*}[t]
\centering
\includegraphics[width=\textwidth]{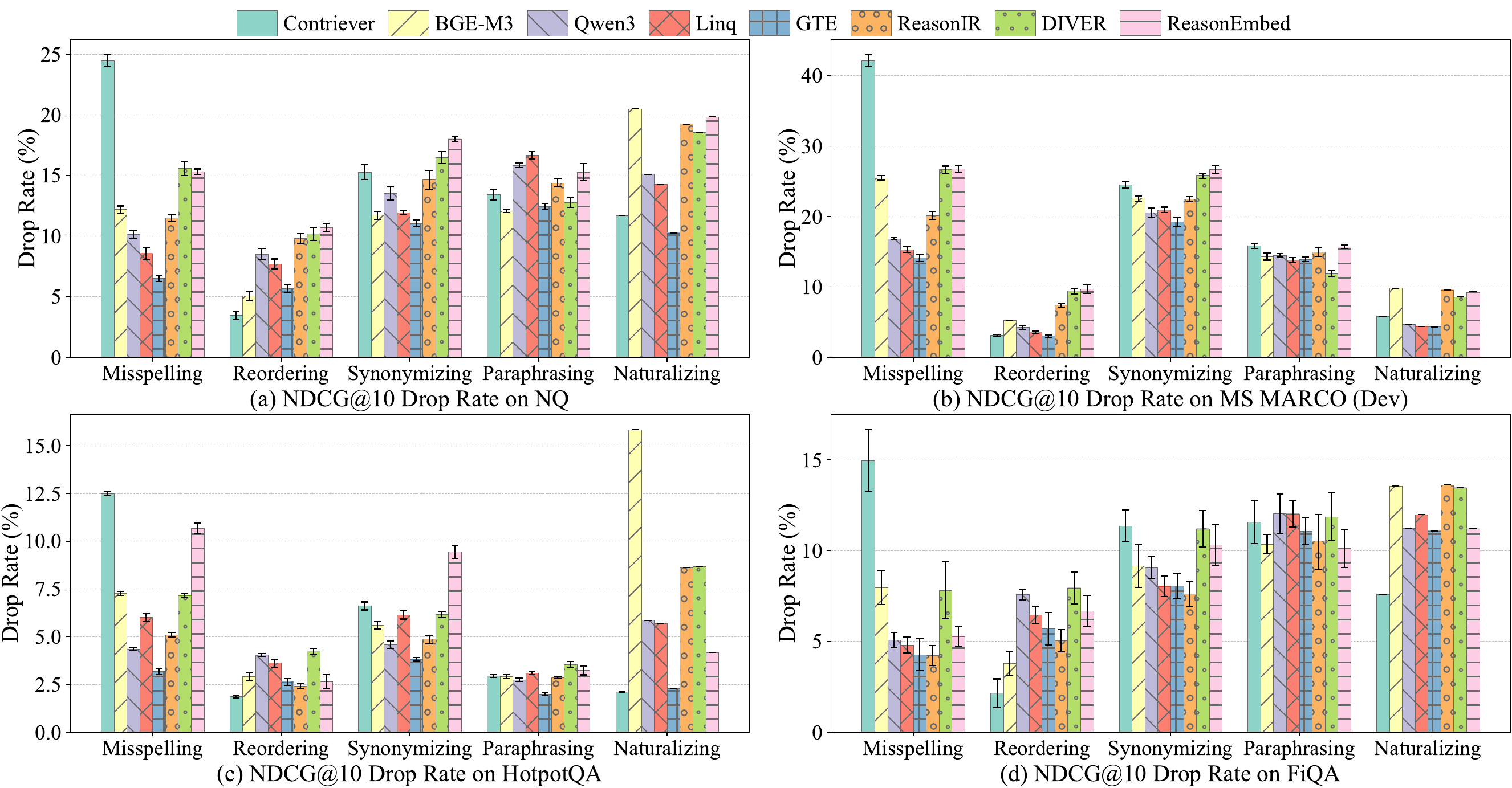}
\caption{Performance~(nDCG@10) drop rate (\%) under five query perturbation types across four datasets. Bars report the mean over 5 seeds, and error bars indicate $\pm$1 standard deviation; lower drop implies higher robustness.}
\label{fig:query_variation_drop_rate_nq}
\end{figure*}

\subsubsection{Results and Discussion}
Figure~\ref{fig:query_variation_drop_rate_nq} reports the performance drop rate under five query perturbation types across four datasets. Several patterns emerge.

\textit{Misspelling.} Misspelling most clearly separates model families on MS~MARCO, where Contriever suffers the largest degradation ($\approx$42\%) and BGE-M3 follows at around 17\%, while the general-purpose instruction-augmented models (Qwen3, Linq, GTE) show substantially smaller drops (6--12\%). Reasoning-oriented models do not share this advantage: DIVER and ReasonEmbed reach roughly 27\% on MS~MARCO, comparable to BGE-M3 and far above their instruction-augmented counterparts. On NQ, the gap between model families is narrower; Contriever still suffers the most ($\approx$24\%), but differences among the remaining models are less pronounced. This pattern is consistent with, but does not by itself establish, the idea that large-scale pretraining on noisy web text improves robustness to character-level noise.

\textit{Synonymizing and Paraphrasing.} Both perturbation types induce substantial drops across all models, with a particularly clear divergence on MS~MARCO: synonymizing causes larger drops (22--27\%) than paraphrasing (13--20\%). A fixed perturbation recipe does not imply equal effective severity, since altering one token affects short keyword-style queries more strongly than longer queries with greater lexical redundancy. This may partly explain why paraphrasing has uniformly low impact on HotpotQA ($\approx$3\%), whose queries are much longer on average (17.6 words) than those of MS~MARCO (6.0). NQ (9.2 words on average), falling between these two extremes in query length, similarly shows an intermediate paraphrasing effect consistent with this length-based explanation. More broadly, synonymizing emerges as a consistently challenging perturbation across architectures, whereas paraphrasing is more dataset-dependent. In both cases, the gap between LLM-based and encoder-only models is much narrower than for misspelling.

\textit{Reordering.} Reordering exposes a partial architectural effect, most clearly on NQ. Contriever in particular is notably robust ($\approx$3\%), while several LLM-based retrievers show higher drop rates; BGE-M3 falls in between and is not consistently more robust than the LLM-based models. This gap is smaller and less consistent on MS~MARCO and HotpotQA. GTE is a notable exception among LLM-based models, achieving drop rates comparable to encoder-only models. One possible explanation is attention directionality: in standard decoder-only models, swapping two words changes the left-context representations of surrounding tokens, which may alter the final pooled embedding more than in bidirectional models. Under this hypothesis, GTE's bidirectional attention design may partly account for its relative stability. We conducted a preliminary ablation by converting Qwen3-Embedding-0.6B, a causal LLM retriever, to bidirectional attention and fine-tuning it on MS~MARCO, but the resulting gains were limited and do not provide strong support for this explanation. One possible reason is that our ablation used only MS~MARCO-scale training, whereas prior work~\cite{eslami2026diffusionpretraineddensecontextualembeddings} suggests that bidirectional conversion may require large-scale continued pretraining to yield substantial improvements. We therefore view attention directionality as a plausible but unverified contributor rather than a confirmed causal mechanism (see Section~\ref{sec:interventions} for a more detailed discussion).

\textit{Naturalizing.} Naturalizing is more dataset-dependent. On MS~MARCO, all models remain relatively robust (3--10\%), consistent with its short keyword-style queries where stopword removal has limited effect. On NQ and HotpotQA, however, BGE-M3 is the most vulnerable model (about 20.5\% and 16\%, respectively), exceeding even Contriever and suggesting a model-specific sensitivity to stopword-stripped queries.

FiQA shows a somewhat different profile: inter-model differences are generally smaller, but synonymizing and paraphrasing still tend to induce larger drops than misspelling. This is broadly consistent with the query-length hypothesis, since FiQA (10.8 words on average) occupies an intermediate regime between MS~MARCO (6.0) and HotpotQA (17.6), although domain-specific financial phrasing may also contribute. The larger error bars further suggest that, under our protocol, FiQA is less discriminative than the other three datasets for separating model differences.

\paragraph{Summary of Query-Side Stability.}
The five perturbation types expose distinct robustness profiles that cannot be collapsed into a single axis. Misspelling is the perturbation where model family differences are most pronounced: the general-purpose instruction-augmented LLM retrievers are clearly more robust than encoder-only models, while reasoning-oriented models do not share this advantage. Reordering is the perturbation where architectural differences may matter, although the pattern is clearest on NQ and weaker on the other datasets. Synonymizing remains challenging across all model families, while paraphrasing is more dataset-dependent and is most disruptive on shorter-query settings such as MS~MARCO and FiQA. Naturalizing is more dataset-dependent, with BGE-M3 showing a marked vulnerability on NQ and HotpotQA. Across all perturbation types, GTE exhibits the most consistently strong stability~(see Section~\ref{sec:interventions} for a discussion of potential explanatory factors). Notably, reasoning-oriented training does not translate into improved query-side stability; this contrasts with the more heterogeneous picture that emerges for document-side attacks in Section~\ref{sec:corpus_poisoning}.

\subsection{Document-Side Stability: Corpus Poisoning Attacks}\label{sec:corpus_poisoning}

In this work, we operationalize document-side stability as robustness to corpus poisoning attacks, which directly manipulate the retrieval corpus by injecting adversarial documents. We focus on this setting because it is the most established retrieval-specific document-side threat model and provides a concrete test of whether retrievers can resist malicious corpus manipulation. This choice does not exhaust the space of document-side perturbations, but it isolates a practically important and security-relevant failure mode. We study corpus poisoning under two attack settings that together cover the main threat scenarios considered in this paper.

\subsubsection{Attack Settings}
Following~\citet{ZhongHWC23_Poisoning,li2025reproducinghotflip}, we inject $|\mathcal{A}| \in \{10, 50\}$ adversarial documents into corpora containing millions of documents and measure the \textbf{Top-20 attack success rate (ASR@20)}, defined as the proportion of test queries for which at least one adversarial document appears in the top-20 results. Adversarial documents are optimized against training queries but evaluated on held-out test queries, thereby assessing whether the attack generalizes beyond the optimization targets. Evaluation is conducted on three representative datasets: NQ, MS~MARCO (Dev), and HotpotQA. All experiments use 5 random seeds (1999, 2016, 2026, 5, 27), and we report mean and standard deviation.

Within this protocol, we consider the following two threat models:

\textbf{(1) White-box attack.} The attacker has full gradient access to the target model (i.e., full access to model weights and architecture) and generates adversarial documents using the optimized HotFlip algorithm~\cite{ZhongHWC23_Poisoning,li2025reproducinghotflip} directly against it. HotFlip operates by iteratively replacing tokens in an adversarial document to maximize its similarity to a set of target query embeddings, so that the injected document is more likely to be retrieved for those queries; first-order gradient approximations are used to guide the discrete token search efficiently. Following \citet{li2025reproducinghotflip}, training queries are partitioned into $|\mathcal{A}|$ clusters via K-means so that each adversarial document is optimized against a semantically coherent subset of queries, improving attack coverage across diverse query intents; one 50-token adversarial document (excluding the task instruction prefix prepended by instruction-tuned models) is generated per cluster. This setting assumes full model access and no defensive measures, representing the worst-case upper bound on attack effectiveness.

\textbf{(2) Direct-transfer black-box attack.} In the broader black-box literature, attacks typically involve first distilling the target model into a surrogate, then applying white-box optimization to the surrogate, and finally transferring the resulting adversarial documents to the target model~\cite{10.1145/3576923_PRADA,Liu0GR0FC23_TARA,Liu0GRFC24_multi_granular,10.1145/3583780.3614793_MCARA}. However, for modern LLM-scale retrievers, surrogate fidelity becomes an additional confounding factor: a weak surrogate can substantially underestimate attackability, making it difficult to disentangle target-model robustness from surrogate quality. Training faithful surrogates for multiple LLM retrievers is also computationally expensive in our setting. We therefore study a narrower black-box setting that avoids surrogate training: following \citet{li2025reproducinghotflip,li2025unsupervised}, we directly inject adversarial documents generated against one model under the white-box setting into every other model's corpus without any model-specific optimization, thereby evaluating cross-architecture transferability under a no-access constraint.

\begin{table}[t]
\centering
\small
\caption{Examples of adversarial documents generated by the white-box HotFlip attack against DIVER. We show short excerpts for readability; the original adversarial documents contain 50 optimized tokens and often include mixed-script and multilingual fragments characteristic of gradient-based adversarial generation.}
\label{tab:adversarial_doc_examples}
\begin{tabularx}{\columnwidth}{lX}
\toprule
\textbf{Dataset} & \textbf{Adversarial Document Excerpt} \\
\midrule
NQ 
    & \texttt{raham-map} [ZH] \texttt{ONA} [ZH] \texttt{MarIES} [ZH] \texttt{HING} \texttt{continent} \texttt{OINST} [TR] \texttt{Country} [ZH] \texttt{Website} [DE] \ldots \\
\addlinespace
MS~MARCO 
    & \texttt{ActualAt} [ZH] \texttt{Ob} [ZH] \texttt{disgr} \texttt{monthlyustral} [ZH] \texttt{months} \texttt{advantages} [ZH] \texttt{incapacexclude} [ZH] \ldots \\
\addlinespace
HotpotQA 
    & [ZH] \texttt{Founded} \texttt{Wit} [ZH] \texttt{Directed} \texttt{helt} \texttt{Franz} \texttt{t~s\'earty} \texttt{heiLTalus} [ZH] \texttt{Education} \ldots \\
\bottomrule
\end{tabularx}
\end{table}

Table~\ref{tab:adversarial_doc_examples} provides examples of adversarial documents generated under the white-box setting.

\begin{table*}[t]
\small
\centering
\caption{
Top-20 attack success rates under corpus poisoning attacks. We first generate $|\mathcal{A}|$ adversarial documents using the training set of each dataset, inject them back into the corpus, and then evaluate whether at least one adversarial document appears in the top 20 retrieved results for held-out test queries. We run 5 random seeds and report the mean and standard deviation in parentheses. Lower is more robust. All ASR@20 values are percentages; the \% symbol is omitted for readability.
}   
\label{tab:results_successrate}
\renewcommand\arraystretch{1.0}
\setlength{\tabcolsep}{1.6mm}{
\begin{tabular}{l|l|rrrrrrrrrrr}
    \thickhline
 \textbf{Dataset} & \textbf{$|\mathcal{A}|$}  &\textbf{Contriever} &  \textbf{BGE-M3}  &  \textbf{Qwen3} &\textbf{Linq} & \textbf{GTE} & \textbf{ReasonIR} &\textbf{DIVER} &  \textbf{ReasonEmbed} \\ 
    \hline

\multirow{2}{*}{{NQ}} & 10  &78.8~\std{4.0}  & 6.3 ~\std{0.6}  & 1.3~\std{0.2} & 4.5~\std{0.5} & \textbf{0.0}~\std{0.0} & 0.8~\std{0.7} & 10.6~\std{1.0} & 5.4~\std{1.9}                \\ 

 &50 & 94.8~\std{1.0} &  21.2~\std{1.1}  & 5.2~\std{0.6} & 14.8~\std{0.7} & \textbf{0.0}~\std{0.0} & 3.8~\std{1.0} & 20.7~\std{0.6} & 11.9~\std{1.0}   
   \\
   \hdashline
\multirow{2}{*}{{MS MARCO}} & 10 &  49.0~\std{8.1}  &  0.6~\std{0.1}  & \textbf{0.0}~\std{0.0} & \textbf{0.0}~\std{0.0} & \textbf{0.0}~\std{0.0} & \textbf{0.0}~\std{0.0} & 5.6~\std{0.4} & 5.4~\std{2.3} \\ 

 &50 &  89.3~\std{1.6} &  6.1~\std{0.3}  & 0.3~\std{0.1} & 0.7~\std{0.2} & \textbf{0.0}~\std{0.0} & 0.1~\std{0.0} & 8.4~\std{0.3} & 10.1~\std{1.9}
   \\
   \hdashline
   \multirow{2}{*}{{HotpotQA}} & 10  & 94.4~\std{0.8} &  22.7~\std{0.7}  & 0.8~\std{0.4} & 15.6~\std{1.1} & \textbf{0.0}~\std{0.0} & 0.1~\std{0.0} & 3.7~\std{0.4 } & 8.7~\std{4.1} \\ 

 &50 &  99.5~\std{0.1} &  55.0~\std{0.9}  & 11.0~\std{1.7} & 43.8~\std{0.6} & \textbf{0.0}~\std{0.0} & 0.7~\std{0.2} & 13.5~\std{0.4} & 24.1~\std{1.6}
   \\
    \hline
    
\end{tabular}
}

\end{table*}

\subsubsection{White-Box Attack Results}
Table~\ref{tab:results_successrate} reports white-box ASR@20, revealing substantial variation in poisoning robustness across retrievers, datasets, and attack budgets.

GTE is the most robust model in this setting, with 0\% ASR across all datasets and both attack budgets; Section~\ref{sec:interventions} investigates the factors underlying this exceptional robustness. ReasonIR is the next most stable retriever, remaining below 4\% even at $|\mathcal{A}|=50$. Qwen3 also shows strong resistance overall, although its ASR rises to 11.0 on HotpotQA at the higher attack budget. Linq is robust on NQ and MS~MARCO, but substantially less stable on HotpotQA, where ASR reaches 43.8 at $|\mathcal{A}|=50$. Rather than establishing a specific mechanism, these results indicate that some LLM-based retrievers are markedly harder to poison under white-box optimization than others.

In stark contrast, Contriever is highly vulnerable: attack success rates rise from 49.0--94.4 at $|\mathcal{A}|=10$ to 89.3--99.5 at $|\mathcal{A}|=50$. As noted by \citet{li2025reproducinghotflip}, this vulnerability is closely related to its dot-product scoring, which allows adversarial documents to exploit L2-norm inflation. BGE-M3 shows intermediate robustness but remains clearly less stable than the strongest LLM-based retrievers, especially as the attack budget increases (e.g., from 22.7 to 55.0 on HotpotQA).

Reasoning-oriented retrievers do not exhibit a consistent robustness advantage. While ReasonIR remains highly resistant, DIVER and ReasonEmbed show moderate vulnerability, with ASR@20 reaching 13.5 and 24.1 on HotpotQA at $|\mathcal{A}|=50$, respectively. This heterogeneity indicates that reasoning-oriented training does not reliably translate into white-box poisoning robustness. Dataset and budget also matter: increasing the attack budget from 10 to 50 substantially amplifies success rates for several models, and HotpotQA is generally the most challenging dataset, especially for Linq, BGE-M3, and ReasonEmbed.

\subsubsection{Direct-Transfer Black-Box Attack Results}
The direct transfer setting yields a uniformly negative result: across all 8 source models, all 8 target models, and all three datasets (using the stronger $|\mathcal{A}|=50$ attack budget), every transfer ASR@20 remains below 1\% over the same 5 random seeds. This remains true even when Contriever---the most vulnerable target under white-box attack, with 89--99\% ASR---is attacked in the no-access setting: adversarial documents crafted against other retrievers still almost never transfer successfully to it. We evaluate all pairwise source-target combinations across the 8 retrievers, but because the resulting transfer ASR@20 values are uniformly near zero, we summarize the outcome in text rather than presenting a dedicated main-text table.

This result is consistent with the broader observation in \citet{li2025reproducinghotflip} that direct transfer attacks are already weak even among encoder-only retrievers trained with broadly similar objectives. In our setting, transfer becomes harder still because the model set spans substantially more heterogeneous architectures and training regimes, including encoder-only baselines, instruction-augmented LLM retrievers, and reasoning-oriented retrievers. The observed near-zero transfer is therefore consistent with weak cross-model transferability, potentially due to differences in architecture, training regime, and embedding geometry.

From a practical perspective, this means that the severe white-box poisoning risk does not directly carry over to the no-access direct-transfer setting studied here. At the same time, this should not be read as implying that all black-box risks are negligible: stronger black-box attacks in the literature typically rely on a faithful surrogate model before white-box optimization, and could in principle approach the white-box upper bound more closely than direct transfer does. Direct transfer therefore appears largely ineffective for modern LLM-based dense retrievers, but surrogate-based black-box attacks remain an important open threat model.

\paragraph{Summary of Document-Side Stability.}
Document-side stability presents a two-part picture. Under white-box corpus poisoning, the strongest LLM-based retrievers---especially GTE, ReasonIR, and Qwen3---show clear robustness advantages over encoder-only baselines. This advantage is not uniform: Linq is substantially less stable on HotpotQA, and reasoning-oriented models such as DIVER and ReasonEmbed exhibit higher vulnerability than the strongest general-purpose models, indicating that reasoning augmentation does not inherently confer adversarial robustness. Under the direct-transfer black-box setting, by contrast, adversarial documents optimized for one retriever rarely remain effective on another, with all transfer ASR@20 values remaining below 1\%.
 
\paragraph{Putting it Together.}
Taken together, the two analyses show that robustness gains from LLM-based retrievers are perturbation-specific rather than uniform. GTE is the most consistently stable model across both query- and document-side evaluations. Semantic-preserving query reformulations remain a substantial source of instability overall: synonymizing is consistently disruptive across datasets, and paraphrasing also induces sizable drops on most datasets despite its much weaker effect on HotpotQA. Reasoning-oriented training, meanwhile, does not confer a consistent stability advantage.

\section{RQ3: What factors are predictive of robustness in LLM-based retrievers?}

To identify potential indicators of robustness, we study three candidate factors using two complementary analyses. For embedding isotropy and spectral norm, we conduct cross-model correlation analyses against the stability metrics reported in RQ2~(Section~\ref{Sec:RQ2_main}). For model size, we perform a within-family scaling analysis in the Qwen3 embedding family. The three factors we examine are: (1) the \textit{geometry of the embedding space}, specifically \textit{isotropy}; (2) \textit{model size scaling}; and (3) \textit{spectral norm} as a proxy for Lipschitz smoothness.

\begin{figure*}[!t]
\centering
\begin{subfigure}[b]{\textwidth}
    \centering
    \includegraphics[width=\textwidth]{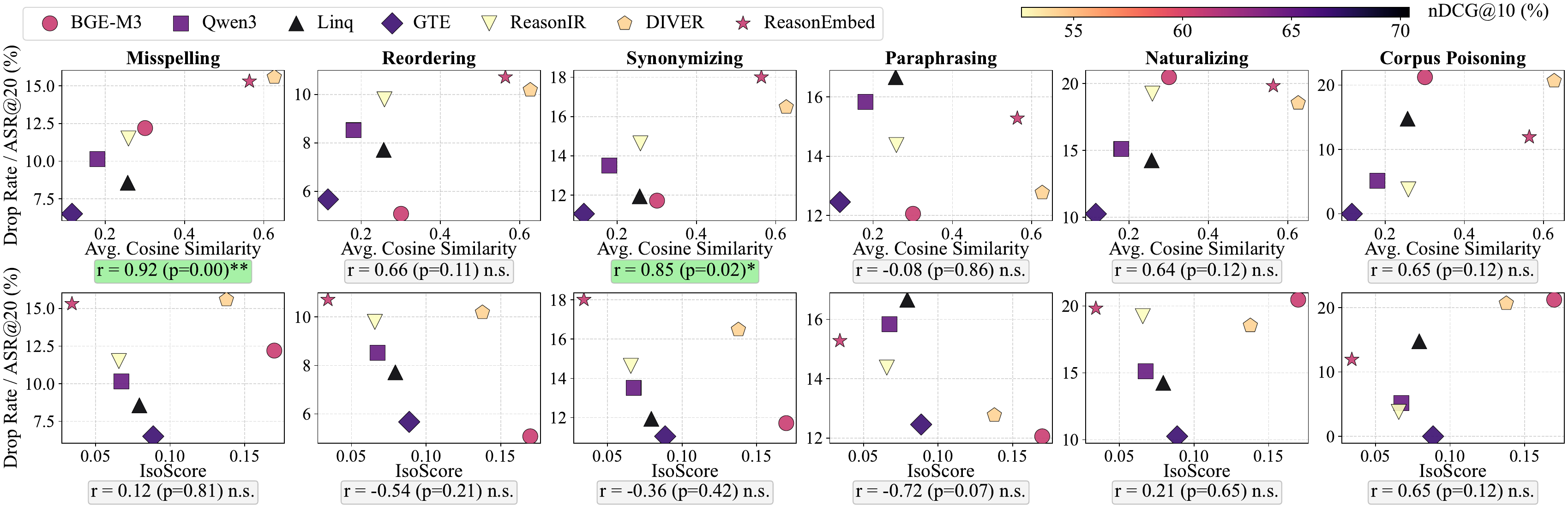}
    \caption{NQ}
    \label{fig:isotropy_query_nq}
\end{subfigure}
\begin{subfigure}[b]{\textwidth}
    \centering
    \includegraphics[width=\textwidth]{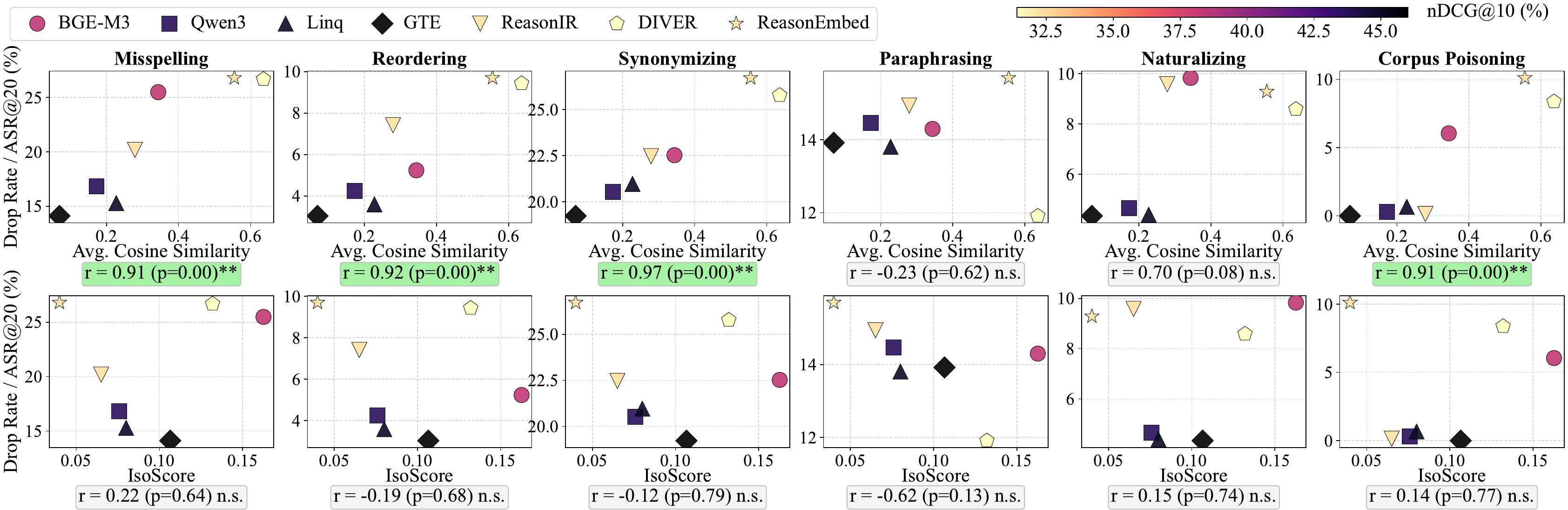}
    \caption{MS~MARCO}
    \label{fig:isotropy_query_msmarco}
\end{subfigure}
\begin{subfigure}[b]{\textwidth}
    \centering
    \includegraphics[width=\textwidth]{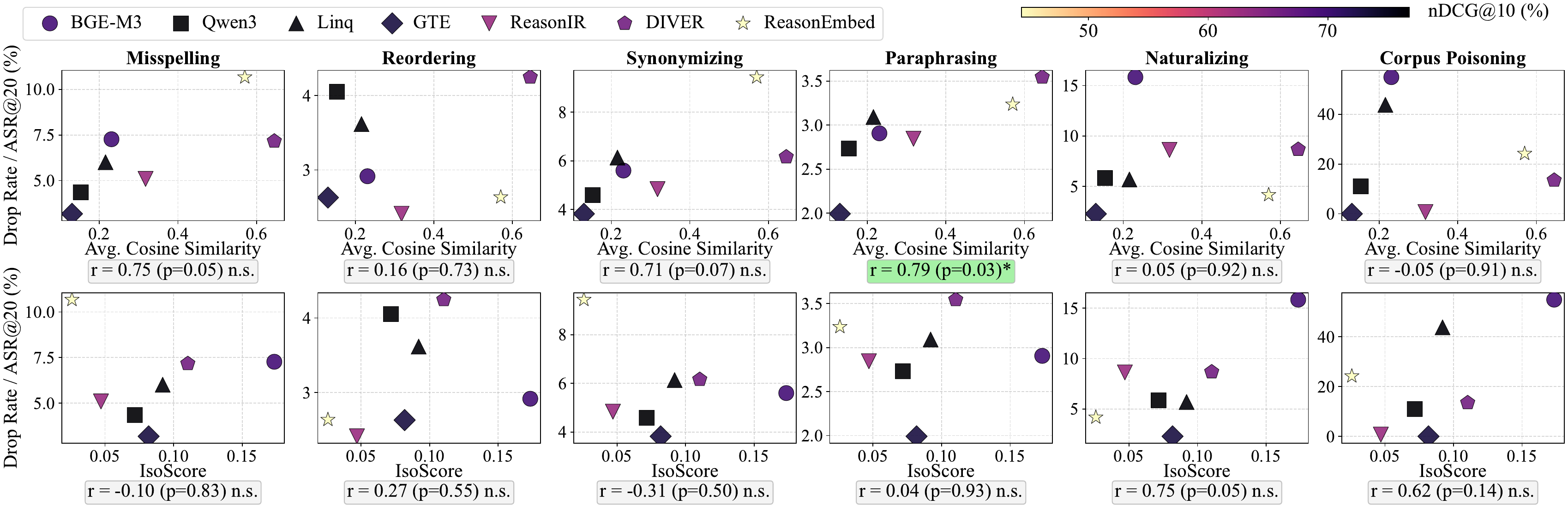}
    \caption{HotpotQA}
    \label{fig:isotropy_query_hotpotqa}
\end{subfigure}
\caption{Pearson correlations between embedding isotropy proxies computed from corpus embeddings and robustness metrics from RQ2~(Section~\ref{Sec:RQ2_main}) across NQ (a), MS~MARCO (b), and HotpotQA (c). Each subfigure shows a $2{\times}6$ grid of scatter plots, where each point represents one retriever. \textit{Top row}: Angular uniformity (avg. cosine similarity) vs. robustness metric; \textit{bottom row}: Variance uniformity (IsoScore) vs. robustness metric. Columns correspond to five query variation types~(nDCG@10 drop rate, \%) and corpus poisoning~(ASR@20, \%). Pearson $r$ and $p$-value are reported in each panel. Box colors: \colorbox{green!30}{$p<0.05$}, \colorbox{black!5}{$p\geq 0.05$}. Point color encodes clean-retrieval nDCG@10 for contextual reference~(see colorbar); marker shapes identify retrievers.}
\label{fig:isotropy_query_additional}
\end{figure*}

\subsection{Embedding Isotropy and Robustness}

Embedding geometry is a broad concept that includes directional concentration, dimensional utilization, norm distribution, and local neighborhood structure. In this subsection, we focus on \textit{isotropy} as a tractable and theoretically motivated geometric descriptor, because prior work has argued that it captures how evenly semantic information is distributed in the embedding space, making it a natural candidate predictor of robustness under perturbation.

\subsubsection{Isotropy Definition and Measurement}
The term \textit{isotropy} in embedding spaces refers to two related but distinct properties:

\textbf{(1) Angular uniformity}: whether embedding vectors are spread across different directions in space, rather than concentrated in a ``narrow cone''~\cite{ethayarajh-2019-contextual,GaoHTQWL19,CaiHB021_Isotropy}. Following prior isotropy analyses, we operationalize this property using \textit{average pairwise cosine similarity}, where lower similarity indicates higher angular uniformity.

\textbf{(2) Variance uniformity}: whether the embedding space is utilized evenly across dimensions, which we measure using \textit{IsoScore}~\cite{RudmanGRE22_IsoScore,RudmanE24_stable_anisotropic}. A score approaching 1 indicates more uniform utilization, while a low score indicates that the embedding space effectively relies on a smaller subset of dimensions.

While both properties have been associated with representation quality~\cite{zhang-etal-2020-revisiting,xiao-etal-2023-isotropy,JungPCKR23,godey-etal-2024-anisotropy}, recent studies suggest that their relationship with downstream performance is more nuanced than previously assumed~\cite{fuster-baggetto-fresno-2022-anisotropy,ait-saada-nadif-2023-anisotropy,RudmanE24_stable_anisotropic}.

\subsubsection{Experimental Setup}
We compute both isotropy metrics for each retrieval model using corpus embeddings from each dataset. We use corpus rather than query embeddings because the corpora provide a much larger and more stable sample for geometric estimation, whereas query sets are comparatively small and vary substantially across datasets. Since each retriever encodes queries and documents into the same embedding space, corpus-side geometry still provides a comparable view of the representation space relevant to retrieval. For \textit{angular uniformity}, we randomly sample 100,000 embedding pairs and compute the average pairwise cosine similarity across these pairs. For \textit{variance uniformity}, we randomly sample 100,000 embeddings and compute IsoScore\footnote{\url{https://github.com/bcbi-edu/p_eickhoff_isoscore/tree/main/IsoScore}} following~\citet{RudmanGRE22_IsoScore}. We exclude BM25, which does not produce dense embeddings, and Contriever, whose scoring is confounded by L2-norm inflation (see Section~\ref{sec:corpus_poisoning}), as including the latter would distort the isotropy--robustness correlations.

\subsubsection{Results and Discussion}\label{sec:rq3_results}
Figure~\ref{fig:isotropy_query_additional} reports Pearson correlations across both isotropy proxies, all perturbation types, and all three datasets. Because this cross-model analysis uses a limited pool of retrievers, we interpret the resulting correlations as small-sample signals rather than definitive estimates. Within that constraint, correlations that are both statistically significant and recurrent across datasets are especially noteworthy.

$\bullet$ \textbf{Angular Uniformity as a Signal of Lexical and Surface-Level Robustness.} High average cosine similarity (low angular uniformity) emerges as the clearest cross-model signal of vulnerability to lexical and other surface-level perturbations. On NQ, we observe significant correlations with \textbf{Misspelling} ($r$=0.92, $p$<0.01) and \textbf{Synonymizing} ($r$=0.85, $p$<0.05). This relationship is even more pronounced on MS~MARCO, where angular uniformity is strongly associated with instability on \textbf{Misspelling} ($r$=0.91), \textbf{Reordering} ($r$=0.92), and \textbf{Synonymizing} ($r$=0.97), all with $p$<0.01. On the same dataset, it also correlates strongly with corpus poisoning ($r$=0.91, $p$<0.01), a pattern consistent with the possibility that concentrated embedding spaces are more vulnerable to attack patterns that depend on exact term alignment. On HotpotQA, Misspelling ($r$=0.75, $p$=0.05) and Synonymizing ($r$=0.71, $p$=0.07) show trends consistent with the other datasets. Notably, Paraphrasing shows a significant correlation on HotpotQA ($r$=0.79, $p$<0.05)---a pattern absent on NQ and MS~MARCO---which we hypothesize may relate to the multi-hop nature of HotpotQA queries. Because this effect does not recur on the other two datasets, we interpret it as an exploratory dataset-specific pattern rather than a stable general effect. Across simpler query settings on NQ and MS~MARCO, angular uniformity shows no consistent association with Paraphrasing, suggesting its primary empirical association is with lexical and surface-level rather than semantic stability. Taken together, these recurrent significant correlations make angular uniformity a promising diagnostic signal, while Section~\ref{sec:interventions} shows that directly regularizing it is not by itself sufficient to improve robustness.

$\bullet$ \textbf{Variance Uniformity: Suggestive but Inconsistent Trends.} Variance uniformity (IsoScore) shows a fragmented relationship with robustness. We find no consistent pattern for query variations: correlations are weak and vary in direction across datasets and perturbation types. For \textbf{Corpus Poisoning}, NQ ($r$=0.65) and HotpotQA ($r$=0.62) both exhibit positive but non-significant trends, whereas MS~MARCO shows only a weak association ($r$=0.14). These results leave open the possibility that higher dimensional utilization is associated with greater poisoning vulnerability in some open-domain QA settings, but the pattern is not yet statistically reliable or cross-dataset consistent. Overall, variance uniformity remains a secondary and currently inconclusive signal relative to angular uniformity.

\begin{figure*}[t]
\centering
\begin{subfigure}[t]{0.49\textwidth}
  \centering
  \includegraphics[width=\linewidth]{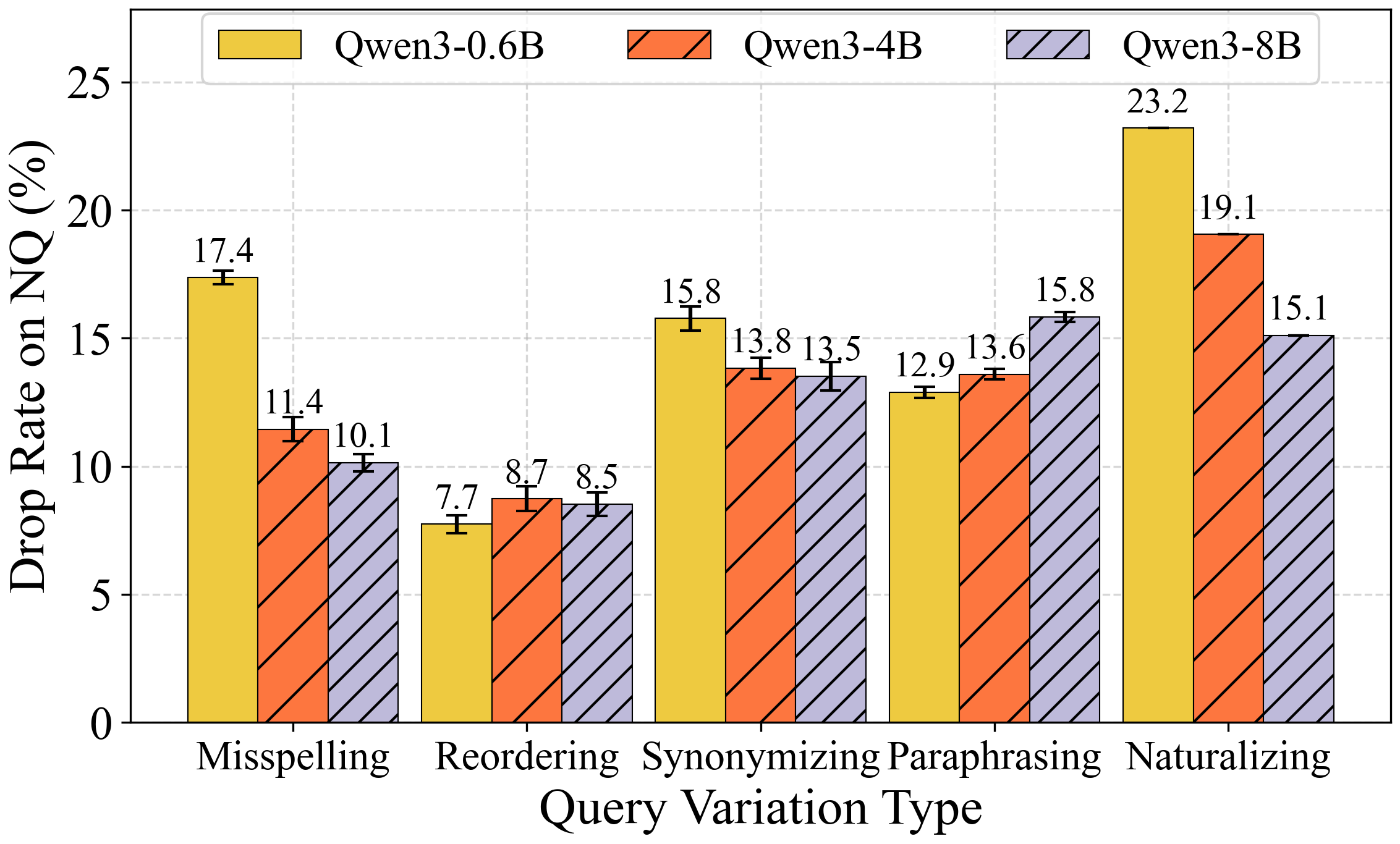}
  \caption{NQ: nDCG@10 drop rate (\%).}
 \label{fig:qwen3_size_queryvar_nq}
\end{subfigure}
\hfill
\begin{subfigure}[t]{0.49\textwidth}
  \centering
  \includegraphics[width=\linewidth]{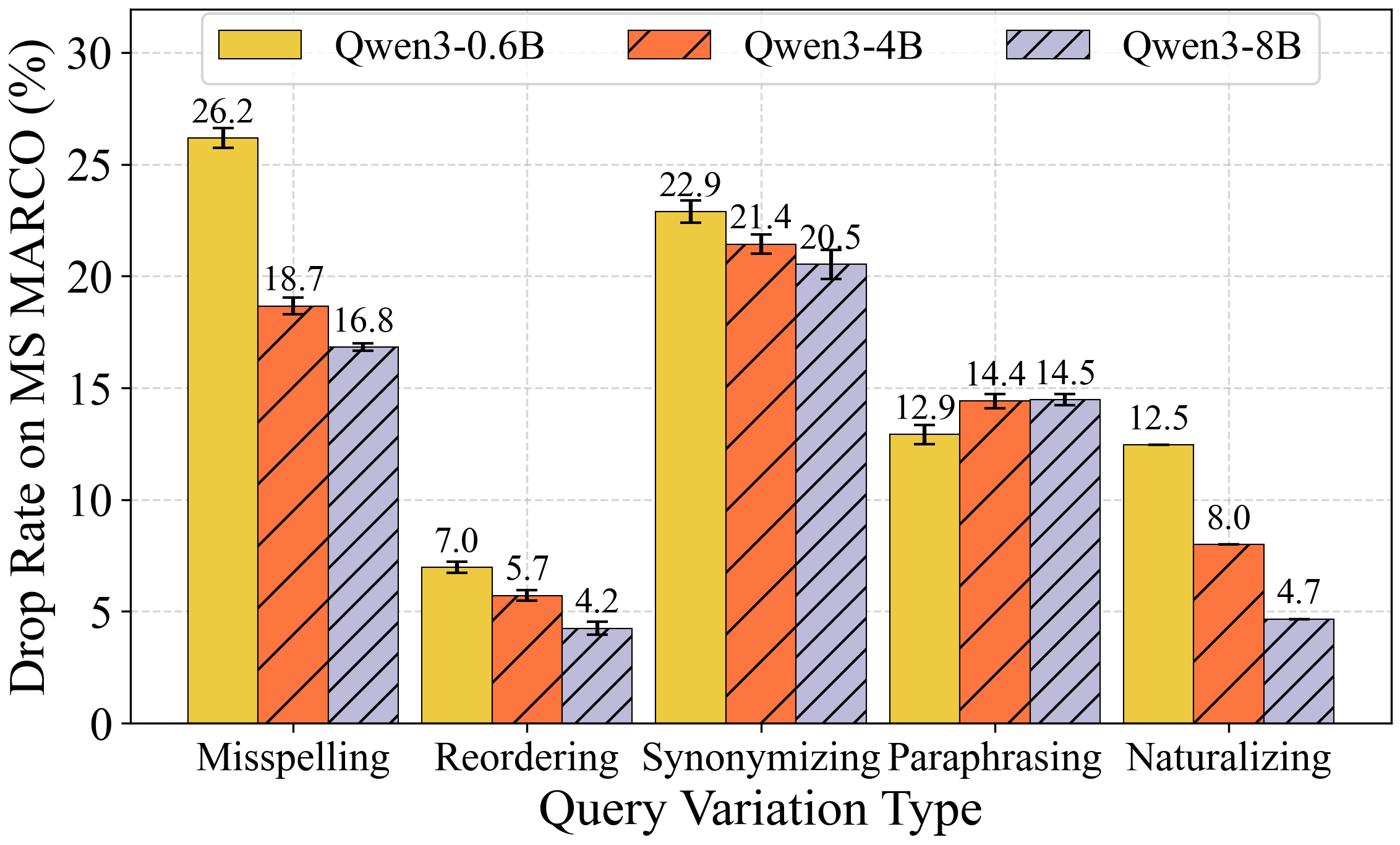}
  \caption{MS~MARCO: nDCG@10 drop rate (\%).}
  \label{fig:qwen3_size_queryvar_msmarco}
\end{subfigure}

\begin{subfigure}[t]{0.49\textwidth}
  \centering
  \includegraphics[width=\linewidth]{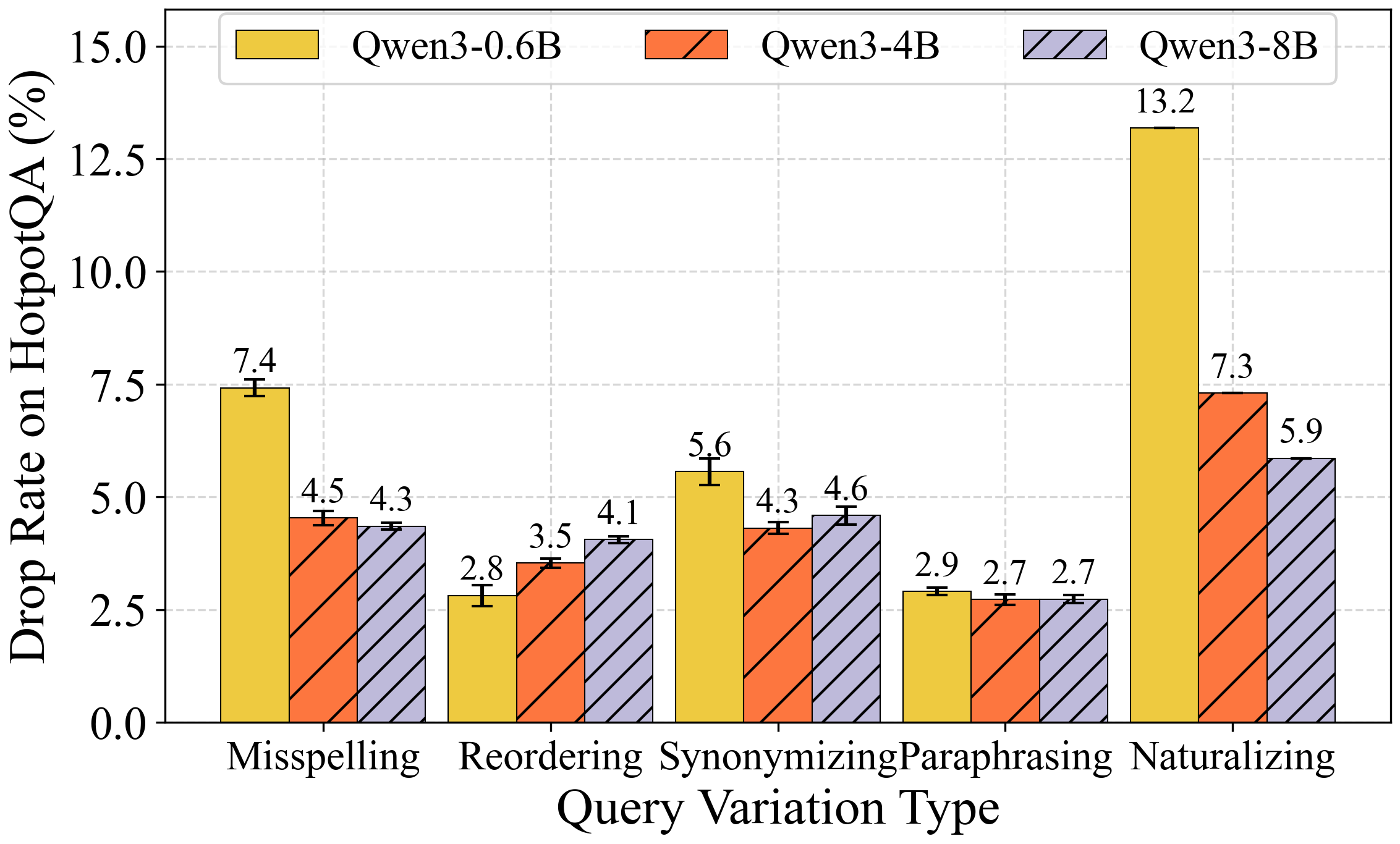}
  \caption{HotpotQA: nDCG@10 drop rate (\%).}
  \label{fig:qwen3_size_queryvar_hotpotqa}
\end{subfigure}
\hfill
\begin{subfigure}[t]{0.49\textwidth}
  \centering
  \includegraphics[width=\linewidth]{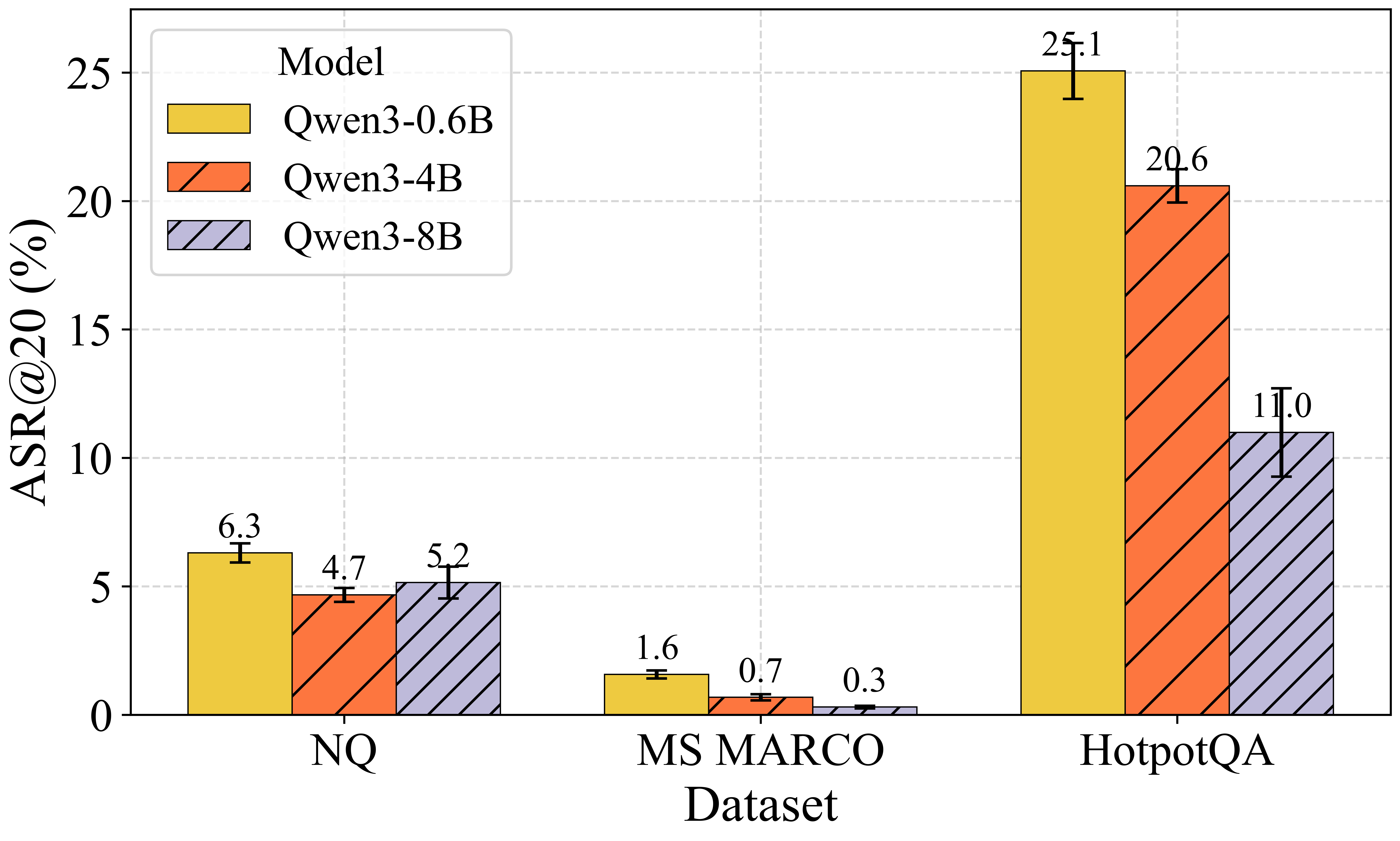}
  \caption{ASR@20 (\%) across datasets.}
  \label{fig:qwen3_size_poison_asr}
\end{subfigure}
\caption{Effect of Qwen3 model size on robustness to query variations and corpus poisoning. (a--c) nDCG@10 drop rate (\%) under five query variation types on NQ, MS~MARCO, and HotpotQA. (d) Corpus-poisoning attack success rate (ASR@20, \%) across datasets. Lower values indicate stronger robustness. Error bars indicate $\pm$1 SD over random seeds.}
\label{fig:qwen3_size_robustness}
\end{figure*}

\subsection{Model Size and Robustness}\label{sec:rq3_model_size}

\subsubsection{Experimental Setup}
Beyond geometric properties, we next ask whether robustness also varies systematically with model scale when the model family is held relatively constant. Prior work by \citet{Liu0GRFC25} suggests that, for encoder-only retrievers, effectiveness and robustness can both improve with size, although the scaling patterns are not always monotonic. It remains unclear whether a similar trend holds for decoder-only embedding models. To reduce architectural confounds, we analyze the \textit{Qwen3 embedding} family, which provides three publicly available sizes (0.6B, 4B, and 8B). Using the same protocol as RQ2~(Section~\ref{Sec:RQ2_main}), we evaluate robustness to both \textbf{query variations}~(nDCG@10 drop rate, \%) and \textbf{corpus poisoning}~(ASR@20, \% at $|\mathcal{A}|{=}50$). Figure~\ref{fig:qwen3_size_robustness} summarizes the results on NQ, MS~MARCO, and HotpotQA.

\subsubsection{Results and Discussion}
For \textbf{query variations} (Fig.~\ref{fig:qwen3_size_robustness}a--c), larger Qwen3 models generally exhibit smaller nDCG@10 drop rates across datasets, indicating improved stability to input perturbations. The trend is most pronounced for \textit{Misspelling}, \textit{Synonymizing}, and \textit{Naturalizing}; for example, on NQ, the Misspelling drop rate decreases from 17.4\% for the 0.6B model to 10.1\% for the 8B model. By contrast, the gains for \textit{Reordering} and \textit{Paraphrasing} are weaker and more dataset-dependent, suggesting that scaling more reliably improves robustness to surface-form perturbations than to structural or semantic reformulations.

For \textbf{corpus poisoning} (Fig.~\ref{fig:qwen3_size_poison_asr}), ASR@20 also tends to decrease with model size. The clearest reduction appears on HotpotQA, where attack success drops from 25.1\% to 11.0\% as model size increases from 0.6B to 8B. The same qualitative pattern holds on MS~MARCO, while NQ shows a mild non-monotonicity in which the 4B model slightly outperforms the 8B model.

These results provide preliminary within-family evidence from the Qwen3 family that larger decoder-only embedding models can be more robust, although the trend is not perfectly monotonic across all settings. Combined with the established improvement in retrieval effectiveness across the Qwen3 family~\cite{qwen3embedding}, they suggest that model size can serve as a useful within-family robustness signal, while broader cross-family validation remains open.

\begin{figure*}[t]
\centering
\begin{subfigure}[b]{\textwidth}
    \centering
    \includegraphics[width=\textwidth]{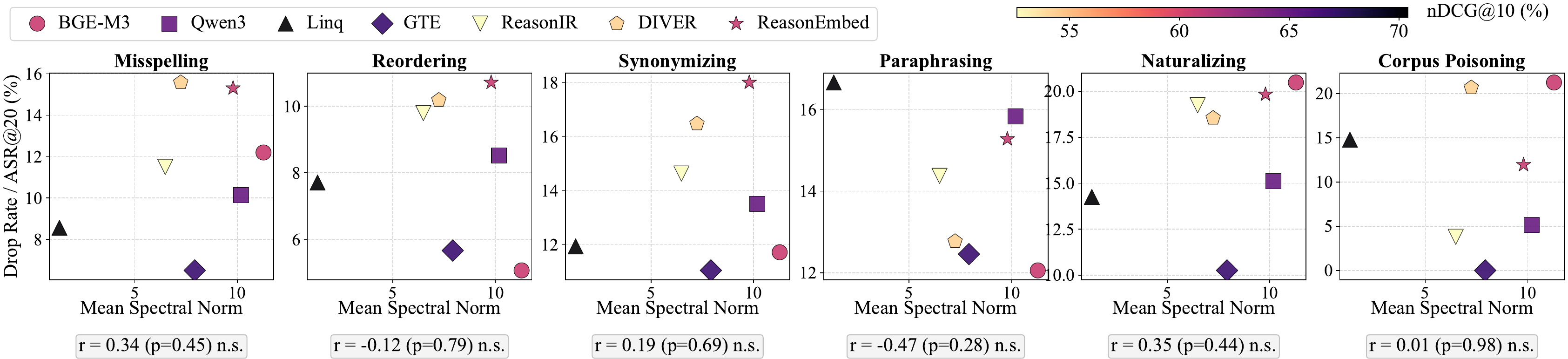}
    \caption{NQ}
    \label{fig:spectral_norm_query_nq}
\end{subfigure}
\\[0.5em]
\begin{subfigure}[b]{\textwidth}
    \centering
    \includegraphics[width=\textwidth]{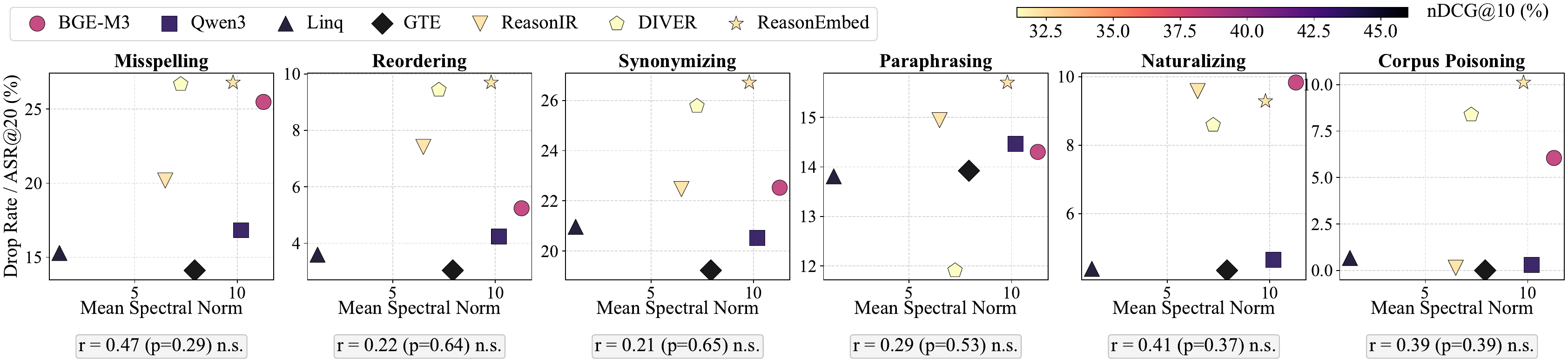}
    \caption{MS~MARCO}
    \label{fig:spectral_norm_query_msmarco}
\end{subfigure}
\\[0.5em]
\begin{subfigure}[b]{\textwidth}
    \centering
    \includegraphics[width=\textwidth]{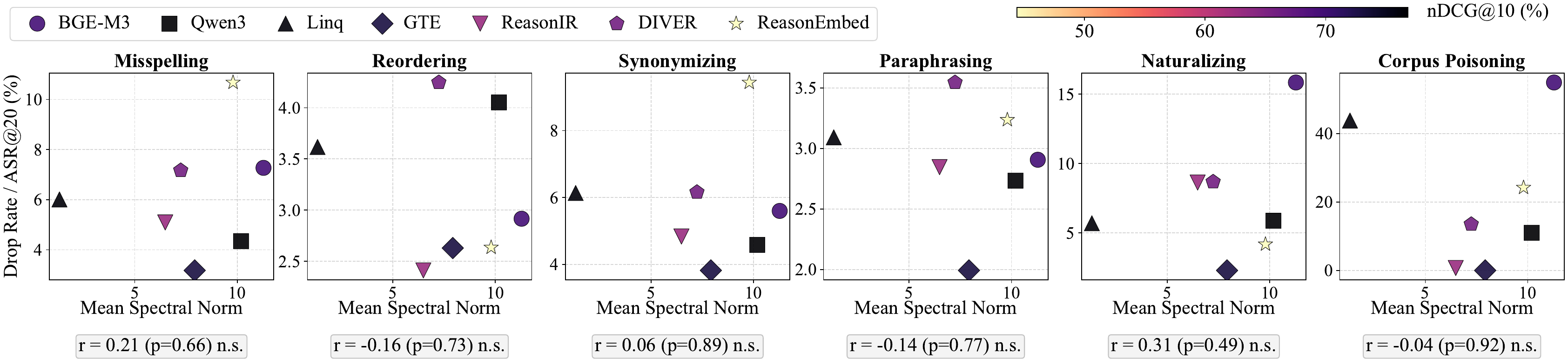}
    \caption{HotpotQA}
    \label{fig:spectral_norm_query_hotpotqa}
\end{subfigure}
\caption{Pearson correlations between mean spectral norm and robustness metrics from RQ2~(Section~\ref{Sec:RQ2_main}) across NQ (a), MS~MARCO (b), and HotpotQA (c). Each subfigure shows six scatter plots, where each point represents one retriever and the x-axis is the mean spectral norm. Columns correspond to five query variation types~(nDCG@10 drop rate, \%) and corpus poisoning~(ASR@20, \%). Pearson $r$ and $p$-value are reported in each panel. Box colors: \colorbox{green!30}{$p<0.05$}, \colorbox{black!5}{$p\geq 0.05$}. Point color encodes clean-retrieval nDCG@10 for contextual reference~(see colorbar); marker shapes identify retrievers.}
\label{fig:spectral_norm_query_variations}
\end{figure*}

\subsection{Spectral Norm and Robustness}
\label{sec:rq3_spectral_norm}

\subsubsection{Experimental Setup}
As a final factor, we examine mean spectral norm as a theoretically motivated proxy for Lipschitz smoothness. As in the isotropy analysis above, we restrict this analysis to the same dense retriever set, excluding BM25 and Contriever for comparability. In continuous settings, lower local Lipschitz sensitivity is often associated with greater robustness~\cite{HeinA17,VirmauxS18,JordanD20,ShiW0KH22}. Because exact local estimates are impractical for large transformer retrievers, we instead use the \textit{mean spectral norm} across layers as an architecture-level approximation. For each retrieval model, we compute the spectral norm of the major transformer weight matrices---including query, key, value, and output projections as well as feed-forward layers---using power iteration, and then average these values across layers.

\subsubsection{Results and Discussion}
Figure~\ref{fig:spectral_norm_query_variations} reports Pearson correlations between mean spectral norm and the robustness metrics from RQ2. Overall, mean spectral norm exhibits no statistically significant correlations with robustness, indicating limited predictive value relative to the other factors examined in this section.

A limited pattern appears for certain query variations. \textbf{Misspelling} shows positive correlations across all three datasets (NQ: $r$=0.35; MS~MARCO: $r$=0.47; HotpotQA: $r$=0.21), and \textbf{Naturalizing} does as well (NQ: $r$=0.35; MS~MARCO: $r$=0.41; HotpotQA: $r$=0.31). These trends suggest that larger spectral norms may be associated with greater sensitivity to lexical noise, although the evidence remains weak and non-significant. By contrast, \textbf{Reordering}, \textbf{Synonymizing}, and \textbf{Paraphrasing} show inconsistent or near-zero correlations.

For \textbf{corpus poisoning}, the results are similarly mixed. MS~MARCO exhibits a modest positive correlation ($r$=0.39), whereas NQ ($r$=0.01) and HotpotQA ($r$=-0.04) show essentially no relationship. Thus, spectral norm does not emerge as a stable predictor of adversarial vulnerability across datasets.

These results suggest that mean spectral norm has only limited, context-dependent value as a robustness indicator. This is not entirely surprising: its theoretical motivation is rooted in continuous perturbation analyses, whereas dense retrieval operates over discrete tokenized inputs with residual connections, layer normalization, and other architectural features that weaken the connection between layer-wise spectral norms and actual retrieval robustness. More precise estimates of local sensitivity in language models remain an open direction for future work.

\paragraph{Putting it Together.}
Taken together, these analyses suggest that embedding isotropy---especially angular uniformity---provides the clearest cross-model signal among the factors examined here. Model size offers preliminary within-family evidence from the Qwen3 family, whereas mean spectral norm shows only weak and context-dependent associations. Overall, the results point to representation geometry as a more informative lens for robustness than the smoothness proxy considered here.

\section{Discussion}\label{sec:discussion}

The RQ1--RQ3 analyses identify several empirical patterns relating model design, generalizability, and robustness outcomes. In this section, we address two follow-up questions that the correlational analyses alone cannot resolve: whether the geometry--robustness associations are causal, and what practical guidance the overall body of evidence offers. \looseness=-1

\subsection{Probing the Correlation--Causation Gap: Intervention Experiments}\label{sec:interventions}

Section~\ref{sec:rq3_results} identifies angular uniformity as a consistent correlate of lexical and surface-level robustness, and raises the question of whether optimizing angular uniformity during training yields robustness gains. To probe this boundary empirically, we conduct two controlled experiments using LoRA fine-tuning of Qwen3-0.6B on MS~MARCO.

\subsubsection{Angular Uniformity Regularization}
We augment the training objective with a cosine penalty that directly regularizes document embeddings to reduce average pairwise cosine similarity. The intervention successfully reshapes the embedding geometry: on FiQA, average cosine similarity decreases from 0.801 to 0.139, while retrieval effectiveness on the same dataset remains unchanged (nDCG@10 = 35.5 in both conditions). However, evaluating the resulting model under the five query-variation perturbation types reveals no consistent robustness improvement. This intervention therefore shows that reducing average cosine similarity via regularization is not, by itself, sufficient to improve query-side stability under our evaluation protocol. We therefore interpret angular uniformity as more useful as a \textit{diagnostic indicator} than as a standalone training signal in this intervention setting.

\subsubsection{Bidirectional Attention Conversion}
Motivated by GTE's bidirectional attention design, we convert Qwen3-0.6B to bidirectional attention under the same controlled setup and fine-tune on MS~MARCO. The resulting model shows no consistent robustness improvement over the causal-attention baseline. As discussed in Section~\ref{Sec:RQ2_main}, prior work~\cite{eslami2026diffusionpretraineddensecontextualembeddings} suggests that bidirectional conversion may require large-scale continued pretraining to yield meaningful gains; our MS~MARCO-scale ablation may be insufficient to reveal such effects. We therefore treat attention directionality as a plausible but unverified contributor, rather than a factor whose effect is established by this ablation.

\subsubsection{GTE as a Reference Point}
GTE is the only retriever achieving 0\% ASR under white-box corpus poisoning and is the most consistently stable model across query-variation types in Section~\ref{Sec:RQ2_main}. Our intervention experiments suggest that neither angular uniformity regularization nor bidirectional attention conversion alone is sufficient to reproduce this level of robustness. Since GTE's training details---data coverage, objectives, and optimization choices---are not publicly documented, definitive attribution is not possible. We therefore treat GTE as a strong empirical reference point under our threat models, and regard identifying the sources of its robustness as an open research question.

\subsection{Implications and Guidelines}\label{sec:implications}

We synthesize the findings from RQ1--RQ3 and the intervention experiments into practical guidelines for evaluation, model selection, and training.

\subsubsection{Evaluation: Cover Multiple Robustness Axes}
The five perturbation types in RQ2 expose qualitatively different failure modes that cannot be captured by a single aggregate robustness score. A model that is robust to character-level noise (Misspelling) may remain vulnerable to semantic reformulations (Paraphrasing), and document-side robustness under corpus poisoning follows a distinct pattern from query-side stability. We therefore recommend that robustness evaluations span at least one character-level perturbation, one semantic reformulation, and one document-side adversarial attack, with results reported separately by perturbation type and dataset. Collapsing across these axes risks masking model-specific failure modes and producing misleading comparisons.

\subsubsection{Model Selection: Robustness Cannot Be Inferred from Scale or Architecture Alone}
Across RQ1 and RQ2, no single coarse attribute---such as parameter count, encoder versus decoder architecture, or reasoning-oriented versus instruction-augmented positioning---reliably predicts either generalizability or robustness. GTE stands out as the strongest overall empirical reference point in our study: it combines high estimated generalizability across heterogeneous retrieval conditions (RQ1) with the most consistently strong stability under both query variations and white-box corpus poisoning (RQ2). Qwen3 shows a similarly broad generalization profile in RQ1 and strong robustness in RQ2, while ReasonIR is especially strong under white-box poisoning despite being less broadly dominant in the factor-conditional generalizability analysis. Taken together, these patterns suggest that deployment-oriented model selection should be tied to the target risk profile: for broad transfer across tasks, query types, and corpus sources, instruction-augmented models such as GTE and Qwen3 are the strongest choices in our evaluation, whereas for settings where white-box corpus poisoning is a primary concern, GTE and ReasonIR provide the strongest empirical robustness among the models we test. Within the Qwen3 family, larger models reduce both query-variation drop rates and corpus-poisoning ASR (Section~\ref{sec:rq3_model_size}), but this within-family scaling trend does not generalize cleanly across architectures. We therefore recommend prioritizing empirically verified behavior under the relevant evaluation conditions over selection based solely on architectural labels or parameter count.

\subsubsection{Training and Loss Design: Geometry Appears More Useful as a Diagnostic than as a Standalone Training Target}
The angular uniformity regularization experiment shows that improving embedding-space geometry does not automatically translate into robustness gains. Practitioners should therefore not treat low average cosine similarity as a standalone training objective for robustness. Instead, angular uniformity is most useful as a \textit{post-hoc diagnostic}: a model exhibiting high average cosine similarity (concentrated embeddings) is likely to be fragile under lexical perturbations, and this observation should motivate investigation into training data diversity and objective design rather than direct geometric penalization. While prior work has begun to address robustness to character-level noise through typo-aware contrastive pretraining~\cite{zhuang2021dealing,CharacterBERT,tasawong-etal-2023-typo,SidiropoulosK24}, our results suggest that synonymizing remains broadly challenging across model families, whereas paraphrasing is more dataset-dependent and is especially disruptive in shorter-query settings such as MS~MARCO and FiQA. Developing training objectives that explicitly target this harder form of perturbation invariance---for example, consistency losses that align representations of paraphrased and original queries---is therefore a higher-priority open direction.

\subsubsection{Threat Model Calibration: White-Box and Transfer Risks Are Asymmetric}
White-box corpus poisoning poses a genuine risk for several models at high attack budgets ($|\mathcal{A}|=50$), with HotpotQA being the most challenging setting. The direct-transfer black-box results, however, show near-zero transfer ASR across all model pairs under the no-access setting studied here (Section~\ref{Sec:RQ2_main}). Practitioners should calibrate their threat models accordingly: white-box robustness should be prioritized when adversarial access to model weights is plausible, whereas under the direct-transfer no-access setting the practical risk appears substantially lower. Surrogate-based black-box attacks, which we do not evaluate, remain an important open threat that could in principle approach white-box attack success rates more closely than direct transfer does.

\section{Conclusion}

This paper presents a systematic analysis of the robustness of LLM-based dense retrievers, covering generalizability across 30 heterogeneous datasets and stability under query-side perturbations and corpus poisoning attacks, with linear mixed-effects models used to control for dataset heterogeneity. LLM-based retrievers exhibit the strongest overall estimated generalizability, while reasoning-oriented models incur a specialization tax: stronger performance on targeted reasoning-heavy settings but weaker transfer across broader evaluation suites. Stability gains are also perturbation-specific. Instruction-augmented models clearly outperform encoder-only baselines on character-level noise, but synonymizing remains broadly challenging across model families, whereas paraphrasing is more dataset-dependent and is especially disruptive in shorter-query settings such as MS~MARCO and FiQA. Under white-box corpus poisoning, GTE achieves perfect resistance (0\% ASR), whereas under the direct-transfer black-box setting studied here, transfer attack success remains below 1\% ASR across all model pairs.

Among the predictors we examine, embedding isotropy---especially angular uniformity---provides the clearest cross-model signal among the factors we study, while model size within the Qwen3 family provides a useful within-family signal and spectral norm shows only weak, context-dependent associations. A controlled intervention further shows that directly regularizing angular uniformity reshapes embedding geometry without improving robustness in our intervention setting, suggesting that angular uniformity is more useful as a post-hoc diagnostic than as a standalone training objective. GTE remains the strongest empirical reference point under our protocol, though the training factors underlying its exceptional robustness are not publicly documented.

Overall, the results show that robustness in LLM-based dense retrieval is multi-dimensional and cannot be inferred from a single architecture choice, model scale, or geometric statistic alone. Robustness evaluation should therefore span multiple perturbation axes, and future work should prioritize training objectives that directly target query reformulation invariance, stronger surrogate-based black-box attack evaluations, and extensions to non-English retrieval settings.\looseness=-1

\section*{Limitations}

This work has several limitations that suggest directions for future research. First, we evaluate a representative but not exhaustive set of LLM-based retrievers; newer models may exhibit different robustness patterns. Second, all experiments are conducted on English datasets, and our findings may not generalize to other languages. Third, for document-side stability, we study gradient-based corpus poisoning under white-box access and a direct-transfer no-access setting, but we do not evaluate stronger surrogate-based black-box attacks or training-time backdoor attacks, both of which may reveal different vulnerabilities. Fourth, our analysis of embedding geometry is primarily correlational, based on a limited cross-model sample, and uses corpus-side embeddings as proxies for the shared retrieval space; the resulting associations should therefore be interpreted as indicative rather than definitive. The intervention experiments in Section~\ref{sec:interventions} also probe only a limited subset of possible causal mechanisms. Fifth, the model-size analysis is confined to the Qwen3 embedding family and should not be interpreted as a general scaling law for LLM-based dense retrievers. Finally, computational constraints limited corpus-poisoning experiments to three datasets; extending to additional domains may uncover further vulnerability patterns.

\section*{Ethical Considerations}
This work studies corpus poisoning and query perturbations in dense retrieval to diagnose failure modes and support the development of more robust retrievers. Our experiments follow established protocols from prior work; we do not introduce new attack algorithms, optimization techniques, or deployment guidance beyond what is already public.

\paragraph{Consistency with intended use and licenses.}
We use publicly released datasets, models, and toolkits strictly for research evaluation, in accordance with their stated licenses and access conditions. When a dataset or model imposes redistribution constraints, we do not re-release the underlying content or any derivative corpus; instead, we release only code, configuration files, and scripts that reproduce our pipeline given that users independently obtain the original resources from their official sources. Our artifacts are intended solely for academic reproducibility and robustness benchmarking.

\paragraph{Responsible release.}
If we release any generated adversarial passages, they are provided only in a controlled form for research benchmarking (e.g., tied to specific evaluation settings and clearly labeled as adversarial), and we include documentation that discourages misuse and clarifies that the artifacts are not intended for deployment against real systems.\looseness=-1

\begin{acks}
This work was partially supported by the China Scholarship Council (202308440220), the LESSEN project (NWA.1389.20.183) of the research program NWA ORC 2020/21 which is financed by the Dutch Research Council (NWO), the PACINO project~(215742) which is financed by the Swiss National Science Foundation (SNSF), and the NSFC project~(No.62372431) which is financed by the National Natural Science Foundation of China. 
All content represents the opinion of the authors, which is not necessarily shared or endorsed by their respective employers and/or sponsors.
\end{acks}

\bibliographystyle{ACM-Reference-Format}
\bibliography{main}







\end{document}